\documentclass[
  aps,
  prb,
  twocolumn,
  superscriptaddress,
  nofootinbib,
  longbibliography
]{revtex4-2}
\usepackage{appendix}
\usepackage{amsmath,amssymb,bm}
\usepackage{graphicx}
\usepackage{dcolumn}
\usepackage{booktabs}
\usepackage{microtype}
\usepackage{capt-of}
\usepackage{xcolor}
\usepackage{booktabs}
\graphicspath{{Figures/}}

\newcommand{\CQ}{C_{\mathrm Q}}
\newcommand{\VQD}{V_{\mathrm{QD}}}
\newcommand{\dVQD}{\delta V_{\mathrm{QD}}}
\newcommand{\meV}{\,\mathrm{meV}}
\newcommand{\ueV}{\,\mu\mathrm{eV}}

\begin{document}

\title{Distinguishing Quantum Capacitance Signatures of a Topological Majorana Wire from a Normal Wire Segment}

\author{Binayyak B. Roy}
\affiliation{Department of Physics and Astronomy, Clemson University, Clemson, SC 29634, USA}

\author{Jay D. Sau}
\affiliation{Condensed Matter Theory Center and Joint Quantum Institute, Department of Physics, University of Maryland, College Park, Maryland 20742, USA }

\author{Sumanta Tewari}
\affiliation{Department of Physics and Astronomy, Clemson University, Clemson, SC 29634, USA}

\begin{abstract}
Majorana zero modes (MZMs) are spatially separated, near-zero-energy excitations expected at the ends of a topological superconducting (TS) wire. A quantum-dot interferometer can probe the quantum capacitance of the TS wire; the location and magnitude of the capacitance resonances provide information about the MZMs, while their magnetic-flux dependence probes coherent coupling to the two Majorana modes. It has been recently shown that a gapless (i.e., $\Delta=0$) wire segment can also exhibit flux-dependent quantum capacitance oscillations through Aharonov-Bohm interference and, with suitable tuning, can reproduce a Majorana-like response. Here we show that the two mechanisms can be distinguished experimentally. In the gapless normal wire segment, the two parity-dependent signals originate from separate energy resonances corresponding to the lack of generic zero-energy states. By contrast, in the topological wire, the low-energy levels with even and odd parity are topologically robust and nearly degenerate in energy; therefore, the two parity branches remain within the same broader resonance region in the quantum-dot potential and persist under independent variations of the dot potential and wire chemical potential. Our results show that the experimentally observed quantum capacitance response should be evaluated not by a particular flux trace at one optimized point in parameter space, but by its persistence over a finite range of independently controlled parameters, including the quantum-dot potential. This parameter space stability, as demonstrated in recent experiments, provides a direct means of ruling out the gapless normal wire segment as the origin of the observed Majorana-like response.
\end{abstract}

\maketitle

\section{Introduction}

Majorana zero modes (MZMs) provide a nonlocal fermionic degree of freedom and constitute a central ingredient in proposals for topological quantum information processing~\cite{Kitaev2003,Nayak2008,wilczek1982quantum,Moore1991,Read2000,Nayak1996}. A promising solid-state realization is based on a spin-orbit coupled semiconductor proximity coupled to an $s$-wave superconductor and subject to a Zeeman field~\cite{PhysRevB.77.220501,sau2010generic,PRLLutchyn2010,PRLOreg2010,sau2010non,TEWARI2010219}. This proposal has motivated extensive experimental activity in semiconductor-superconductor (SM-SC) heterostructures~\cite{ScienceMourik2012,Deng2012,NatPhysDas2012,Rokhinson_2012,churchill2013superconductor,finck2013anomalous,NatureAlbrecht2016,deng2016majorana,zhang2017ballistic,chen2017experimental,PRLNichele2017,albrecht2017transport,sherman2017normal,o2018hybridization,shen2018parity,vaitiekenas2018selective,Yu_2021,zhang2021,PhysRevB.107.245423,MicrosoftAzureQuantumNature2025,MicrosoftQuantumPb2026,MicrosoftQuantumNatureReply2026}. Nevertheless, establishing the topological origin of an observed low energy state remains a central challenge. Most experimental probes are sensitive to local tunneling amplitudes, or local spectral weight rather than to a bulk topological property. Consequently, finite-size effects, spatial inhomogeneity, and disorder can produce robust low energy states that reproduce tunneling conductance signatures conventionally associated with spatially separated MZMs~\cite{PRBKells2012,PhysRevB.96.075161,PRBMoore2018,Moore2018,SciPostVuik2019,PRRPan2020,PRBZeng2022,DasSarmaSauStanescuPRB2023,RoyJaiswalStanescuTewariPRB2024}.

Interferometric transport of electrons across a Majorana wire through electron teleportation~\cite{Tewari_2008,PRLFu2010,PRBSauSwingleTewari2015,NatCommunWhiticar2020} can, in principle, provide a direct probe of the nonlocal fermionic degree of freedom associated with the topological phase. The underlying connection is that the topological invariant of a Majorana wire can be related to a change in the ground state fermion parity under the insertion of a superconducting flux quantum~\cite{PUKitaev2001}. On the other hand, the Byers-Yang theorem~\cite{PRLByersYang1961}, together with the superconducting condensate, constrains the equilibrium response of a sufficiently large conventional superconductor to be $h/2e$ periodic. This constraint can be circumvented by using Coulomb interactions to make the interferometric response depend on the charging energy and fermion parity~\cite{PRLFu2010,PRBSauSwingleTewari2015,PRBVijayFu2016,PRBPlugge2017}. In these proposals, Coulomb blockade fixes or controls the relevant parity sector, allowing the nonlocal single electron transfer associated with Majorana modes to appear in the interferometric response.

This interferometric connection between fermion parity and nonlocal Majorana transport also underlies the recently developed quantum capacitance approach. Flux-dependent quantum capacitance measurements provide a parity-sensitive probe of a nanowire-quantum-dot (QD) interferometer~\cite{MicrosoftAzureQuantumNature2025,PRBSau2025,StanescuTewariPRB2026}. The quantum capacitance is determined by the curvature of the parity-resolved energy with respect to the dot potential, while the magnetic flux controls the relative contributions of the two interferometer arms and the dot potential tunes the QD into resonance with the low energy wire states. Recent experiments demonstrated single-shot random telegraph switching between two flux-dependent capacitance branches, interpreted as changes in the fermion parity associated with the occupation of a low energy state in the wire ~\cite{MicrosoftAzureQuantumNature2025,MicrosoftQuantumPb2026}. Comparable occupation of the two branches is consistent with a low energy fermionic state, while the flux-dependent crossings of quantum capacitance, with the even and odd parity sectors shifted by $h/2e$, provide the interferometric behavior expected when this state is associated with Majorana modes localized near the two ends of the wire. Importantly, the response was measured as a function of both flux and dot detuning potential and was observed across multiple dot transitions, with transition-dependent visibility and phase. The measurements were also performed in a wire-plunger, which controls the chemical potential, and magnetic field regime independently selected using the topological-gap protocol~\cite{PhysRevB.107.245423}. 

The above measurements constitute an important advance in parity-sensitive interferometry. However, the observation of two flux-dependent branches at a single operating point, defined by the dot potential, does not by itself determine whether the underlying wire state is topological. A distinct counterexample was recently presented by Pinchenkova \textit{et al.}, who showed that a fully non-superconducting finite wire segment can generate $h/e$-periodic quantum capacitance oscillations with an approximate half-period displacement between two particle number sectors~\cite{PinchenkovaKozinHunenbergerLossKlinovaja2026}. We refer to this realization as the \emph{gapless-wire} interferometer, where ``gapless" refers to non-superconducting state $\Delta=0$. The description of its quantum capacitance in even and odd particle number sectors requires two appropriate wire energy levels, denoted by $E_{\tau}$ and $E_g$, together with a QD level positioned between them. The responses identified with the two parity sectors originate from the one- and two-particle sectors and are associated with resonant coupling to the two distinct wire levels. When the detunings of the QD level from $E_{\tau}$ and $E_g$ are comparable, the two capacitance branches can acquire similar amplitudes and closely reproduce the flux dependence expected from a pair of MZMs. Their results therefore establish that flux periodicity and the relative displacement of two capacitance branches are not, by themselves, sufficient to establish either superconductivity or topology.

%It was further noted that this construction requires the wire chemical potential to place the midpoint of the two relevant levels sufficiently close to zero energy and that, in the presence of disorder, the response can be recovered by appropriately retuning the chemical potential and dot potential. 

The existence of this counterexample leaves open a more experimentally relevant question: does the Majorana-like response in the gapless wire remain stable when the experimentally controlled parameters are varied? The distinction is important because the two branches responsible for the two values of quantum capacitance have different microscopic origins in the gapless and topological systems. In the gapless wire, they are tied to two separate single-particle resonances, so moving the QD potential toward one resonance necessarily moves it away from the other. By contrast, in the topological wire, the two branches correspond to the two occupations of the low energy fermionic mode formed from the end MZMs. When the Majorana splitting is small, their capacitance resonances remain centered within nearly the same dot potential region. The two systems may therefore produce similar flux traces at a specially selected operating point in the QD potential while exhibiting parametrically different stability under experimentally accessible detuning.

In this work, we distinguish these mechanisms by calculating the parity-resolved quantum capacitance as a function of magnetic flux, dot potential $V_{\rm QD}$, wire chemical potential $\mu$, and Zeeman energy $\Gamma$. We first reproduce the clean gapless-wire benchmark of Ref.~\cite{PinchenkovaKozinHunenbergerLossKlinovaja2026} and compare it with a clean topological nanowire. The Majorana-like response in the gapless wire occurs when the QD potential is tuned between the two values corresponding to the narrow parity resonances separated by $35.0\ueV$. The separation of the two parity resonances of the dot potential is consistent with the $33.3\ueV$ separation of the two relevant wire levels, $E_\tau$ and $E_g$. By contrast, for the topological Majorana wire, the corresponding parity resonances in the QD potential are separated by only $5.25\ueV$, consistent with $4E_M\sim5.35\ueV$, where $E_M$ is the Majorana splitting energy. Since the even and odd parity resonances as a function of $\VQD$ are so close to each other, the Majorana-like response in the topological wire occupy a substantially broader region in gate voltage controlling the QD potential. Here, by ``Majorana-like response" we refer to bimodal equal intensity oscillation of capacitance as a function of flux $\Phi$ with $\Delta C_Q = |C_Q^{(\rm even)} - C_Q^{(\rm even)}|$ having a periodicity of $h/2e$. The parity resolved quantum capacitance response has an inferred $h/e$ periodicity and flux shifted from each other by a period of $h/2e$.

We next examine how these signals evolve under experimentally relevant parameter variations other than the QD potential. Changing the wire chemical potential, controlled by a plunger gate translates the two separate dot potential resonances, corresponding to the two parity sector branches of the quantum capacitance response of the gapless normal wire, relative to the QD operating point, so retaining comparable parity branches requires a compensating adjustment of $V_{\rm QD}$. By contrast, at the topological point considered here, the two capacitance profiles remain centered within approximately the same gate region throughout the interval $-50\ueV\leq\delta\mu\leq50\ueV$. In addition, we check whether bringing the energy levels of the gapless normal wire closer together by tuning an experimental parameter enhances the robustness of the Majorana-like capacitance response as a function of $\VQD$. To do this, we construct a configuration in the gapless wire in which the energy levels $E_\tau$ and $E_g$ are nearly degenerate and, as a result, the two capacitance resonances are separated by only $3\ueV$, comparable to the topological wire. Although this additional tuning improves the similarity of the two branches of capacitance at one operating point in $\VQD$, we find that the resultant response remains strongly sensitive to variations of $V_{\rm QD}$ and Zeeman field $\Gamma$. These distinctions are directly accessible experimentally as the width, displacement, and relative visibility of the two branches under dot-plunger, wire plunger, and magnetic field sweeps.

Finite temperature occupation and charge fluctuations impose further experimentally observable constraints. At $T=50~\mathrm{mK}$, equilibrium Boltzmann weighting strongly favors one particle number sector in the gapless wire segment, whereas both parity sectors in the topological wire retain equal appreciable weight. If transitions between the sectors occur on a timescale shorter than or comparable to the measurement acquisition time, this imbalance appears directly as a strong asymmetry between the populations of the two capacitance branches. Reducing the energy level separation in the gapless wire segment can restore comparable thermal weights at the optimized point, but a detuning of the Zeeman energy by only $15\ueV$ from the optimized value $\Gamma_0=0.500~\mathrm{meV}$ strongly redistributes the statistical weight between the two sectors. In addition, quasistatic charge fluctuations with an energy scale up to $15\ueV$, comparable to the detuning from the optimized $\VQD$ operating point, in the gapless wire segment, to either resonance, disrupts the balanced and approximately half-period-shifted response. We confirm that these conclusions remain valid even in the presence of disorder.

Our results show that, for the gapless normal wire segment, the bimodal, equal-intensity oscillation of quantum capacitance, with $\Delta C_Q$ having a period of approximately $h/2e$, is restricted to a narrow interval of the QD potential $\VQD$. This effect arises from the finite energy separation (approximately $33.3~\ueV$ for the relevant parameters used in our work, as well as in Ref~\cite{PinchenkovaKozinHunenbergerLossKlinovaja2026}) between the parity resonances in the normal wire as a function of the QD potential. By contrast, the similar Majorana-like capacitance response for the topological wire remains robust over a broad range of parameters, including the quantum dot potential, $\VQD$. Moreover, in the clean case shown in Fig. 2, the capacitance signal in the normal-wire scenario is approximately an order of magnitude smaller than in the topological case at the value where the even- and odd-parity signals are equal. In the disordered case, we find the discrepancy to be even larger, as shown in Fig. 12 in the appendix. This amplitude difference provides an additional quantitative distinction between the two scenarios. 
%experimentally observed quantum capacitance response is distinguished not by a particular flux trace at one optimized point in parameter space, but by its persistence over a finite range of independently controlled parameters, including the quantum-dot potential. 
Since the quantum capacitance response in the recent experiments ~\cite{MicrosoftAzureQuantumNature2025,MicrosoftQuantumPb2026} occurs with $h/2e$ periodicity of $\Delta C_Q$ across many quantum dot transitions for several values of $\VQD$, we conclude that this parameter space stability provides a direct means of ruling out the gapless normal wire segment as the origin of the observed Majorana-like capacitance response.

%Our results therefore do not claim that parameter-space stability alone establishes topological superconductivity. Instead, they convert the specific two-level gapless construction into a directly testable and strongly constrained alternative explanation. The relevant experimental criterion is not merely the observation of periodic, parity-shifted capacitance oscillations at one optimized value of $V_{\rm QD}$, but the persistence and appreciable visibility of both branches under variations of $V_{\rm QD}$, $\mu$, and $\Gamma$, together with independent evidence for a gapped topological superconducting regime. The gapless construction remains a valid counterexample to an interpretation based on a single flux trace, but it cannot account for a response that remains stable throughout a finite multidimensional operating region without repeated fine tuning.

The remainder of this paper is organized as follows. In Sec.~II, we introduce the nanowire-QD model, define the parity-resolved quantum capacitance response, and describe the gapless-wire benchmark. In Sec.~III, we compare the clean gapless and topological interferometers, examine their stability under variations of the dot potential, wire chemical potential, and Zeeman energy, and analyze Finite temperature parity visibility and the more strongly tuned gapless wire configuration. Section~IV summarizes our main results and conclusions. Appendix~A examines the sensitivity of the quantum capacitance response in the gapless wire segment to quasistatic charge fluctuations, Appendix~B considers the role of competing particle number sectors, and Appendix~C presents the representative calculation including Rashba spin-orbit coupling and correlated disorder.

\section{Model and quantum capacitance response}
\label{sec:model}

We describe the proximitized and gapless interferometers within the same finite nanowire-QD geometry and evaluate their parity-resolved quantum capacitance response on equal footing. We first introduce the microscopic model and numerical response function, and then specialize to the gapless-wire benchmark of Ref.~\cite{PinchenkovaKozinHunenbergerLossKlinovaja2026}.

\subsection{Nanowire-QD interferometer}

We consider a spinful one dimensional semiconductor nanowire with $N=300$ lattice sites and lattice spacing $a=10~\mathrm{nm}$, corresponding to a wire length $L=3~\mu\mathrm{m}$. The semiconductor Hamiltonian is
\begin{align}
H_{\mathrm{SM}}={}&
\sum_i c_i^\dagger
\left[
(2t-\mu+V_i)\sigma_0+\Gamma\sigma_x
\right]c_i
\nonumber\\
&+
\sum_i
\left[
c_{i+1}^\dagger
\left(
-t\sigma_0+\frac{i\alpha}{2}\sigma_y
\right)c_i\right]
+\mathrm{H.c.},
\label{eq:HSM}
\end{align}
where $t$ is the nearest-neighbor hopping, $\mu$ is the chemical potential, $\alpha$ is the Rashba coupling, $\Gamma$ is the Zeeman energy, and $V_i$ is an onsite disorder potential. The clean calculations use $V_i=0$. For the disordered calculations discussed in the Appendix, $V_i$ is a smooth zero mean random potential with root-mean-square strength $V_0=1.2~\mathrm{meV}$ and correlation length $\xi_{\mathrm{dis}}=20~\mathrm{nm}$~\cite{DasSarmaSauStanescuPRB2023,RoyJaiswalStanescuTewariPRB2024}.

\begin{figure*}[t]
\centering
\includegraphics[width=0.93\textwidth]{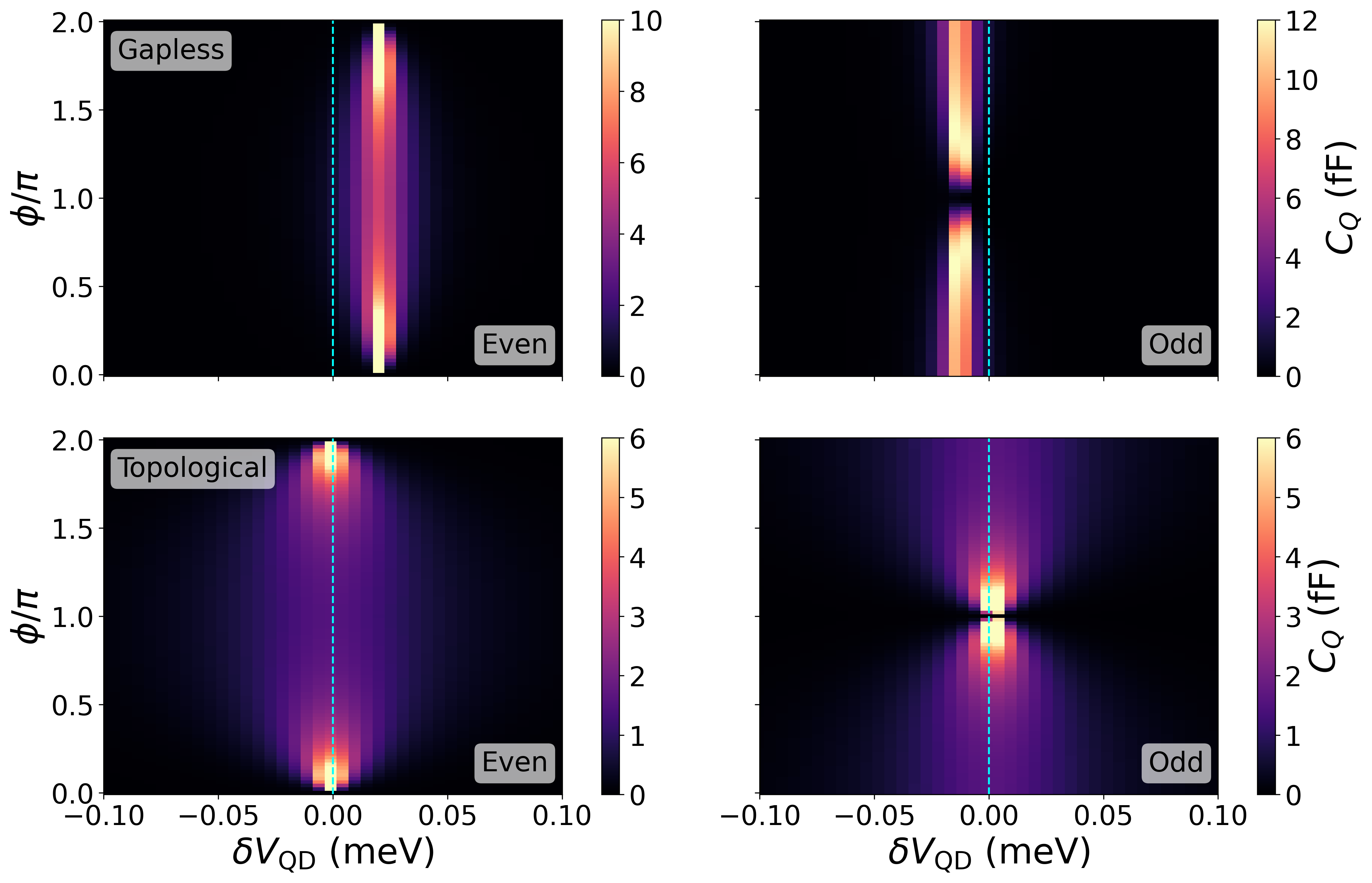}
\caption{\label{fig:clean_maps}\textbf{Parity-resolved capacitance maps for the clean gapless and topological interferometers.} The upper panels show $|\CQ|$ in the even and odd sectors of the clean gapless wire at $\mu=-0.48\meV$ and $\Gamma=0.50\meV$. The even and odd parity resonances are centered near $\dVQD\simeq+19.4\ueV$ and $-13.9\ueV$, respectively, relative to $\VQD^*=0.585\meV$. The lower panels show the corresponding topological responses at $\mu=0.50\meV$ and $\Gamma=0.65\meV$, using $\VQD^*=0.686\meV$. The Majorana-like response in the gapless wire segment occurs between two narrow parity resonances, whereas the topological branches occupy the same broader gate region. Cyan dashed lines mark $\dVQD=0$.}
\end{figure*}

For the proximitized wire, the parent superconductor is integrated out through the local self-energy
\begin{equation}
\Sigma_{\mathrm{SC}}(\omega)
=
-\gamma
\frac{
\omega+\Delta_0(\Gamma)\sigma_y\tau_y
}{
\sqrt{\Delta_0^2(\Gamma)-\omega^2}
},
\label{eq:selfenergy}
\end{equation}
where $\gamma$ is the semiconductor-superconductor coupling and $\tau_y$ acts in particle-hole space. In the static approximation, the semiconductor parameters are renormalized by $Z=\Delta_0/(\Delta_0+\gamma)$ and the induced pairing is $\Delta_{\mathrm{ind}}=\gamma\Delta_0/(\gamma+\Delta_0)$~\cite{StanescuTewariPRB2026,RoySauTewari2026}. Equivalently, the low energy Hamiltonian takes the form $H_{\mathrm{eff}}=Z H_{\mathrm{SM}}+H_\Delta$, with the proximitised superconductivity contribution as
\begin{equation}
H_\Delta
=
\sum_i
\left(
\Delta_{\mathrm{ind}}\,
c_{i\uparrow}^{\dagger}c_{i\downarrow}^{\dagger}
+\mathrm{H.c.}
\right).
\label{eq:HDelta}
\end{equation}
We use
\begin{equation}
\Delta_0(\Gamma)
=
\Delta_0(0)
\left[
1-\left(\frac{\Gamma}{\Gamma_c}\right)^{5/2}
\right],
\qquad
\Gamma_c=1.25~\mathrm{meV},
\label{eq:parentgap}
\end{equation}
with $\Delta_0(0)=0.3~\mathrm{meV}$ and $\gamma=0.2~\mathrm{meV}$. Unless stated otherwise, the proximitized calculations use $t=16.56~\mathrm{meV}$, $\alpha=1.4~\mathrm{meV}$, and dimensionless end couplings $w_L=w_R=0.025$.

The wire is coupled at both ends to a single spinful QD with a dot potential $V_{\mathrm{QD}}$. Following the QD description of Ref.~\cite{PinchenkovaKozinHunenbergerLossKlinovaja2026}, we take
\begin{equation}
H_{\mathrm{QD}}
=
c_{\mathrm{QD}}^\dagger
\left(
V_{\mathrm{QD}}\sigma_0+\Gamma\sigma_x
\right)
c_{\mathrm{QD}},
\label{eq:HQD}
\end{equation}
where $c_{\mathrm{QD}}=(c_{\mathrm{QD}\uparrow},c_{\mathrm{QD}\downarrow})^T$. The same Zeeman term is also included on the dot, while superconducting pairing is restricted to the wire. The parameters $w_L$ and $w_R$ multiply the hopping matrix on the two dot-wire links, and the magnetic flux is introduced as a relative phase on one link. We define
\begin{equation}
\phi=\pi\frac{\Phi}{\Phi_0},
\qquad
\Phi_0=\frac{h}{2e},
\label{eq:fluxconvention}
\end{equation}
so that a $2\pi$ period in $\phi$ corresponds to an $h/e$ flux period.

The parity-resolved quantum capacitance is determined by the curvature of the corresponding many-body energy,
\begin{equation}
C_Q^{(p)}
=
-e^2\alpha_g^2
\frac{\partial^2 E_p}
{\partial V_{\mathrm{QD}}^2},
\label{eq:CQcurvature}
\end{equation}
and is evaluated numerically from the equivalent zero-frequency Kubo response~\cite{PRBSau2025,StanescuTewariPRB2026, RoySauTewari2026}. We set the lever arm $\alpha_g=1$ and use $e^2=0.1602~\mathrm{fF\,meV}$ and, unless noted otherwise, $\eta=0.002~\mathrm{meV}$. The parity contrast used below is
\begin{equation}
\Delta C_Q
=
\left|
C_Q^{(\mathrm{even})}
-
C_Q^{(\mathrm{odd})}
\right|.
\label{eq:DeltaCQ}
\end{equation}

\subsection{Gapless-wire benchmark}

To realize the gapless interferometer proposed in Ref.~\cite{PinchenkovaKozinHunenbergerLossKlinovaja2026}, we set $\alpha=\Delta_0=\gamma=0$ while retaining a finite Zeeman energy. We use $t=102~\mathrm{meV}$ and end couplings $w_L=w_R=0.020$, following the gapless normal wire construction of Ref.~\cite{PinchenkovaKozinHunenbergerLossKlinovaja2026}. In this limit, particle number is conserved and the low energy interferometric response can be described in terms of two relevant single-particle wire levels, $E_\tau$ and $E_g$, coupled to a spin-resolved QD level with energy $E_D$. The Majorana-like response arises when the QD energy level is positioned between the two wire energy levels.

Within the projected one- and two-particle blocks, the two capacitance branches are~\cite{PinchenkovaKozinHunenbergerLossKlinovaja2026}
\begin{equation}
C_Q^j(\phi)
=
\frac{
2e^2\alpha_g^2|j(\phi)|^2
}{
\left[
(E_D-E_j)^2+4|j(\phi)|^2
\right]^{3/2}
},
\qquad
j\in\{\tau,g\},
\label{eq:gaplessCQ}
\end{equation}
where $j(\phi)$ defines the coupling of the individual energy levels $E_\tau$ and $E_g$ to the QD. The $E_\tau$ branch belongs to the $N=1$ odd sector, while the $E_g$ branch belongs to the $N=2$ even sector. Thus, unlike the topological case, the two parity-resolved responses in the gapless wire segment originate from resonant coupling to two single-particle wire energy levels.

Equation~\eqref{eq:gaplessCQ} provides the analytical static-limit description of the effective low energy model\cite{PinchenkovaKozinHunenbergerLossKlinovaja2026}. In the calculations presented below, however, the capacitance response in the gapless normal wire is evaluated from the full tight-binding Hamiltonian using the same zero-frequency Kubo formulation introduced above, with $\eta=0.002~\mathrm{meV}$; the curvature expression is recovered in the static $\eta\rightarrow0$ limit.

In the main-text comparison, we restrict the particle-number sectors to $N=1$ and $N=2$ to give the gapless wire segment its most favorable configuration for reproducing the Majorana-like response. In principle, these sectors can be isolated through a suitable charging-energy hierarchy. However, if other particle-number sectors do enter the low-energy spectrum, the $N=0$ or $N=3$ states can become energetically preferred and suppress one of the parity-resolved capacitance responses. The unrestricted-sector calculation is presented in Appendix~\ref{app:number_sectors}.

Having established a common microscopic framework for the optimized gapless wire segment and the topological interferometer, we now compare their capacitance responses as experimentally accessible parameters are varied. In the following section, we test the robustness of the quantum capacitance response in the gapless and topological interferometers under such parameter variations.

\section{Results}

We begin by comparing the quantum capacitance profiles of the clean gapless and topological interferometers and then examine how they behave under the variation of experimentally accessible parameters. This allows us to distinguish a Majorana-like trace obtained at an optimized operating point from a response that persists over a finite region of control parameter space.

\subsection{Clean wire comparison}
\label{sec:clean}

For the clean gapless wire segment calculation, we use $\mu=-0.480\meV$, $\Gamma=0.500\meV$, and the reference dot potential $\VQD^*=0.585\meV$, together with the hopping and end coupling parameters specified in Sec.~\ref{sec:model}. The two relevant isolated wire energy levels are $E_\tau=-8.89\ueV$ and $E_g=24.44\ueV$, corresponding to dot potential resonances at $\VQD^{(\tau)}\simeq0.571\meV$ and $\VQD^{(g)}\simeq0.604\meV$. Thus, the operating gate selected to reproduce the Majorana-like response lies between the two resonances $\VQD^\tau<\VQD<\VQD^g$. The topological reference is evaluated at $\mu=0.50\meV$, $\Gamma=0.65\meV$, and $\VQD^*=0.686\meV$, with the remaining parameters given in Sec.~\ref{sec:model}. For both systems, we define
\begin{equation}
\dVQD=\VQD-\VQD^*.
\label{eq:VQD-detuning}
\end{equation}
Here, $V_{\rm QD}^*$ denotes the reference dot potential. For the topological interferometer, it is chosen near the center of the common resonance region. For the gapless interferometer, it is chosen approximately midway between the two distinct dot potential resonances, where the two parity responses can acquire comparable amplitudes and produce a Majorana-like response when the system parameters are appropriately tuned.

\begin{figure}[h]
\centering
\includegraphics[width=\columnwidth]{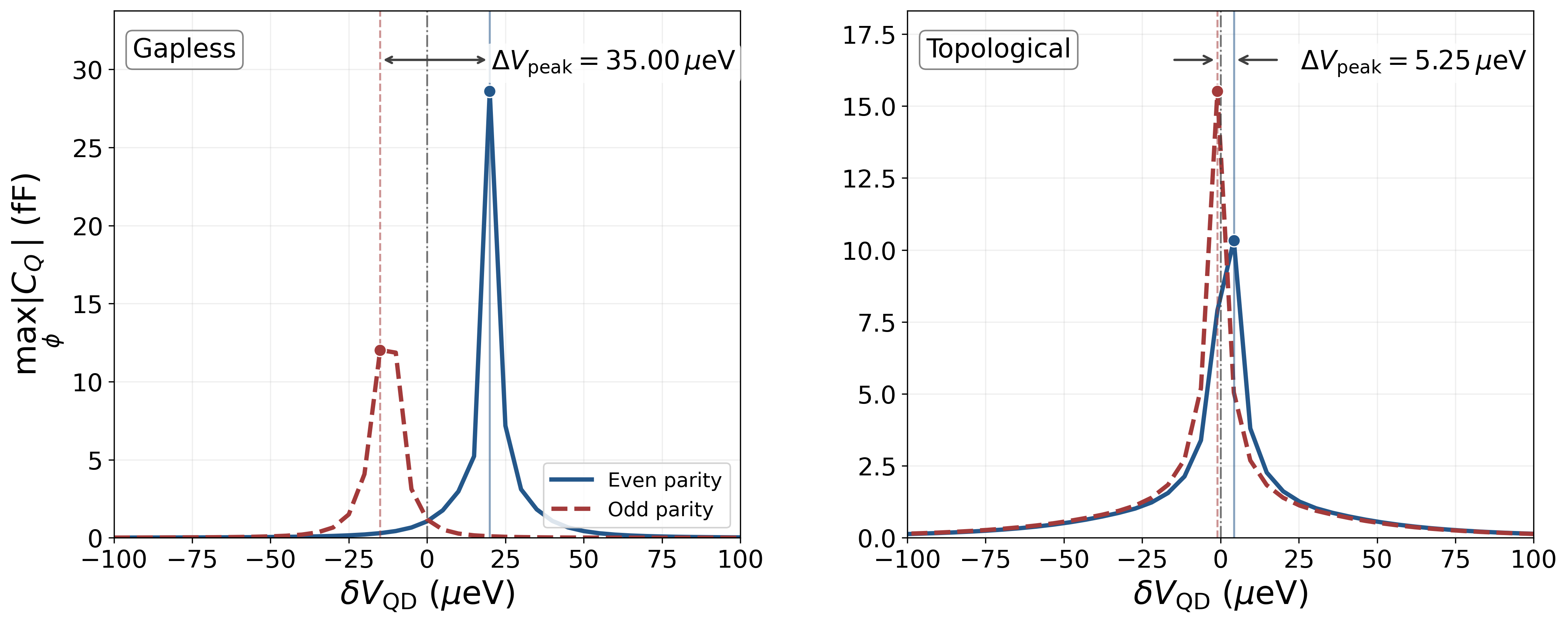}
\caption{\label{fig:clean_peaks}\textbf{Flux-maximized parity-resolved capacitance in the clean systems.} Maximum of $|\CQ|$ over $\phi$ as a function of $\dVQD$ for the gapless (left) and topological (right) wires. Solid blue and dashed red curves denote even and odd parity, respectively. The gapless wire maxima lie near the separate $\tau$ and $g$ resonances and are separated by $\sim35.00\ueV$, whereas the topological wire maxima remain within the same gate region and are separated by $5.25\ueV$. Colored vertical lines mark the maxima, the black dash-dotted line marks $\dVQD=0$.}
\end{figure}

Fig.~\ref{fig:clean_maps} shows the parity-resolved capacitance as a function of $\dVQD$ and the flux phase $\phi$. In the gapless wire, the even response is concentrated near the $\VQD$ value at which the QD couples resonantly to the $E_g$ energy level, while the odd response is centered near the $\VQD$ value at which the QD couples resonantly to the $E_\tau$ energy level. Their positions relative to the selected reference QD potential are $\dVQD^{(g)}\simeq+19.4\ueV$ and $\dVQD^{(\tau)}\simeq-13.9\ueV$. At $\dVQD=0$, the dot is therefore resonant with neither wire energy level. Instead, the tails of the two narrow responses acquire comparable amplitudes and generate the approximately $\pi$-shifted, $2\pi$-periodic parity traces identified in Ref.~\cite{PinchenkovaKozinHunenbergerLossKlinovaja2026}. The Majorana-like response is therefore realized at a specifically selected QD potential between two distinct dot potential resonances for the two parity sectors in the quantum capacitance response of the gapless normal wire.

\begin{figure*}[t]
\centering
\includegraphics[width=0.97\linewidth]{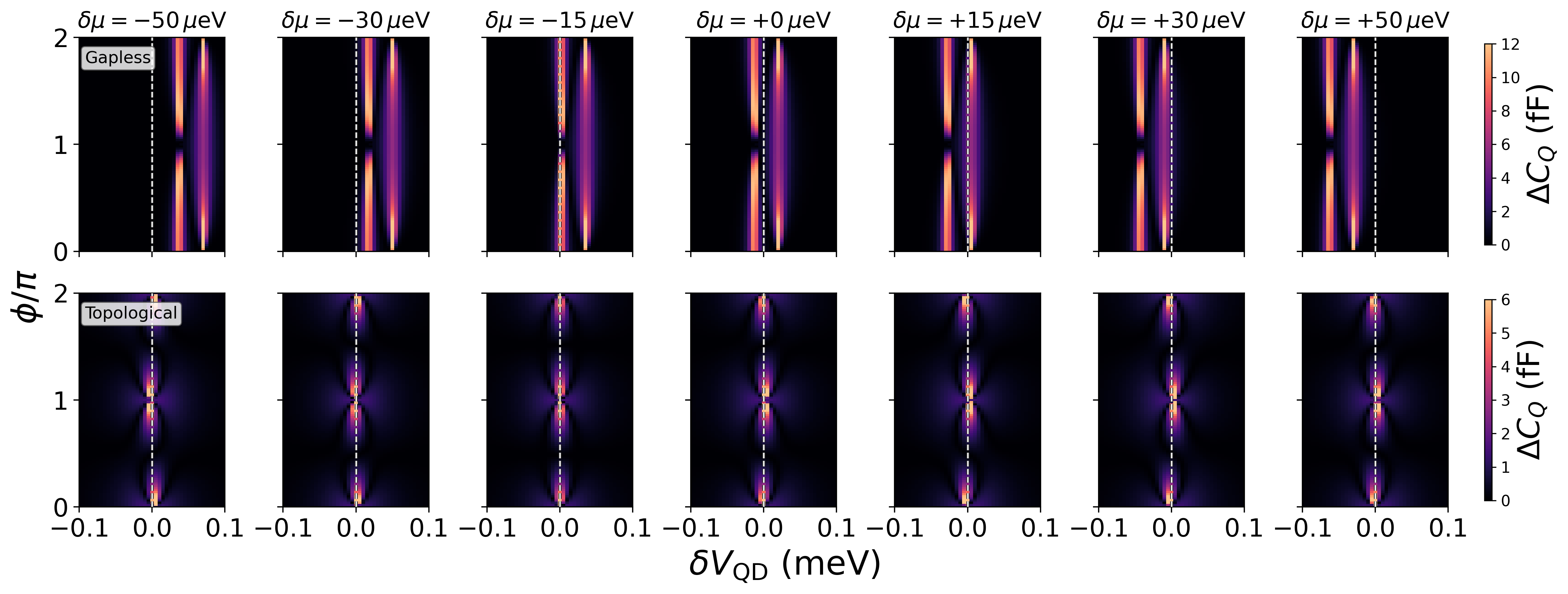}
\caption{\label{fig:mu_detuning}\textbf{Stability under chemical-potential detuning.} Parity contrast $\Delta C_Q$ as a function of $\dVQD$ and $\phi$ for $\delta\mu=(-50,-30,-15,0,15,30,50)\ueV$. The upper row shows the gapless wire with $\mu_0=-0.48\meV$ and $\Gamma=0.50\meV$; its two narrow resonances move across the unchanged operating gate as $\mu$ is varied. The lower row shows the topological wire with $\mu_0=0.50\meV$ and $\Gamma=0.65\meV$; the response remains within the same gate region.}
\end{figure*}

The lower panels of Fig.~\ref{fig:clean_maps} show a qualitatively different gate dependence. In the topological wire, the two branches correspond to the different occupations of the same low-energy fermionic mode formed from the end MZMs~\cite{PRBSau2025,StanescuTewariPRB2026}. The separation between the dot potential resonances in the topological case is of order $\sim4E_M$, where $E_M$ is the Majorana splitting energy. Since $E_M$ is generally small for well-separated MZMs in a robust topological regime, the centers of the two resonances almost overlap. In addition, both capacitance responses extend over a broad interval of $\VQD$. This broad distribution shows that the topological Majorana response persists as the dot potential is varied. Consequently, the parity-shifted flux dependence does not require positioning the dot between two well-separated resonances and remains visible when the dot gate is detuned from the reference value.

The QD potential dependence is quantified in Fig.~\ref{fig:clean_peaks}, where
\begin{equation}
C_{Q,\mathrm{peak}}^{(p)}(\dVQD)
=
\max_\phi
\left|
C_Q^{(p)}(\dVQD,\phi)
\right|.
\label{eq:flux-max-capacitance}
\end{equation}
For the gapless wire, the odd- and even-parity maxima occur near $\dVQD\sim-15.0\ueV$ and $\sim20.0\ueV$, respectively, giving a separation $\Delta V_{\mathrm{peak}}^{\mathrm{G}}=35.00\ueV$. This value closely follows the separation between the two energy levels $E_g-E_\tau=33.33\ueV$, with the remaining difference arising from dot-wire hybridization and the maximization over flux. Importantly, the selected operating point $\dVQD=0$ coincides with neither maximum and instead lies between the two narrow resonances.

In the topological wire, the two maxima are separated by only $\Delta V_{\mathrm{peak}}=5.25\ueV$, close to the corresponding Majorana scale $4E_M=5.35\ueV$. More importantly, the widths of the two topological responses overlap strongly, providing a common gate interval over which both parity branches remain appreciable. Fig.~\ref{fig:clean_peaks} therefore makes explicit the distinction between a response obtained by balancing the tails of two separate narrow dot resonances associated with two distinct energy levels and a response supported by two parity branches within the same QD-potential resonance region, corresponding to different occupations of the same low-energy fermionic mode.

The magnitude of the parity contrast provides another distinction between the two mechanisms. At the intermediate dot potential where the gapless wire produces a Majorana-like response, $\Delta C_Q^{\max}$ is of order $1~\mathrm{fF}$, while the topological response reaches approximately $10~\mathrm{fF}$ near its common resonance. This difference is not simply a consequence of the particular benchmark parameters. In the gapless wire, the dot potential resonances for the two parity branches are visibly separated, reflecting the energy difference between the wire levels $E_\tau$ and $E_g$. The Majorana-like flux dependence appears only in the region between these resonances, where their tails have comparable amplitudes. Moving $V_{\rm QD}$ toward either resonance can produce a larger capacitance signal, but one parity branch then dominates and the comparable, half-period shifted flux responses are lost. Thus, within the two-level approximation of the gapless wire segment, the parity contrast cannot be increased by approaching a resonance without simultaneously destroying the Majorana-like feature. The topological wire does not have this restriction because both parity branches lie within the same broad resonance and retain their characteristic flux dependence near its maximum. The relevant comparison is therefore between the largest parity contrast compatible with a Majorana-like response in the gapless wire and the parity contrast at the QD-potential resonance of the topological wire. Microscopic parameters can change these numerical values, but they do not remove this tradeoff as long as the dot potential resonances of the two parity sectors of the quantum capacitance response, for the gapless wire, remain tied to two distinct wire energy levels.

A direct test of the stability implied by these different QD potential resonance structures is obtained by varying the wire chemical potential while keeping the reference dot potential fixed. Fig.~\ref{fig:mu_detuning} shows $\Delta C_Q(\dVQD,\phi)$ for $\delta\mu=(-50,-30,-15,0,15,30,50)\ueV$. Within each row, $\Gamma$ and the reference dot potential $\VQD^*$ remain unchanged, and the white dashed line marks the fixed reference dot potential. The gapless wire calculations continue to be restricted to the $N=1$ and $N=2$ sectors, thereby giving the gapless normal wire its most favorable realization throughout this comparison.

In the gapless wire, varying $\mu$ shifts the two narrow $\VQD$ resonances associated with the different parity sectors along the $\dVQD$ axis. Increasing $\mu$ moves both resonances toward negative $\dVQD$, while decreasing $\mu$ moves them toward positive $\dVQD$. As a result, the comparable amplitudes of the two parity responses at the original QD potential are rapidly lost, even though the individual resonances remain well defined. The Majorana-like configuration can, in principle, be recovered by compensating for the change in $\mu$ through a corresponding retuning of $\VQD$. However, the need for such fine-tuning reflects the restricted region of parameter space over which the two resonance tails have comparable amplitude leading to a Majorana-like response.

For the topological wire, the parity contrast remains centered near $\dVQD=0$ throughout the displayed $100\ueV$ range of chemical-potential detuning. Both parity branches continue to occupy approximately the same dot potential region without requiring a compensating adjustment of $\VQD$, showing that the topological capacitance response remains stable under the variations in $\mu$ considered here. This behavior can be tested experimentally. In Ref.~\cite{MicrosoftAzureQuantumNature2025}, the flux-dependent capacitance response was measured as a function of the QD plunger voltage across several successive dot transitions [see Fig.~3(i) and Supplementary Fig.~S11 of Ref.~\cite{MicrosoftAzureQuantumNature2025}]. The wire chemical potential and QD potential can also be controlled independently during device tuning~\cite{MicrosoftAzureQuantumNature2025,PhysRevB.107.245423}. The gapless wire segment therefore requires simultaneous fine-tuning of these independent controls to preserve its Majorana-like response. In contrast, the topological response remains stable over a finite region of the same parameter space, providing a direct experimental distinction between the two cases.

\subsection{Finite temperature visibility and fine-tuned response in gapless wire segment}
\label{sec:clean_visibility}

The preceding comparison considers the capacitance at $T=0$ within fixed fermion-parity sectors. We now account for the energy difference between the even- and odd-parity sectors and determine whether both sectors retain appreciable Boltzmann weight at finite temperature. We define
\begin{align}
p_p(\phi)&=
\frac{e^{-E_p(\phi)/(k_{\mathrm B}T)}}
{\sum_{p'}e^{-E_{p'}(\phi)/(k_{\mathrm B}T)}},
\nonumber\\
P(C_Q\mid\phi)&=
\sum_p\frac{p_p(\phi)}{\sqrt{2\pi}\sigma_C}
\exp\left[-\frac{\left(C_Q-C_Q^{(p)}(\phi)\right)^2}
{2\sigma_C^2}\right],
\label{eq:capacitance_distribution}
\end{align}
where $E_p$ is the energy assigned to parity sector $p$, and $\sigma_C$ represents the capacitance-readout resolution. Equation~\eqref{eq:capacitance_distribution} describes the equilibrium limit in which repeated parity switching and relaxation allow both sectors to be sampled according to their Boltzmann weights. Individual measurements performed on timescales shorter than the parity dwell time need not themselves sample this equilibrium distribution, however, over repeated switching events between the two parity sectors the separation in energy between the two corresponding energy levels is expected to bias their relative populations whenever relaxation is governed approximately by thermal equilibrium. We use $T=50~\mathrm{mK}$, corresponding to $k_{\mathrm B}T=4.31\ueV$, and a readout noise broadening of $\sigma_C=0.105~\mathrm{fF}$~\cite{MicrosoftAzureQuantumNature2025}. For the clean system comparison, the probabilities are obtained from the uncoupled-wire energies since we are working in the weak effective dot-wire coupling regime here, where coupling to the QD only weakly perturbs the energies of the relevant wire states while generating the measurable capacitance response.

\begin{figure}[t]
\centering
\includegraphics[width=\columnwidth]
{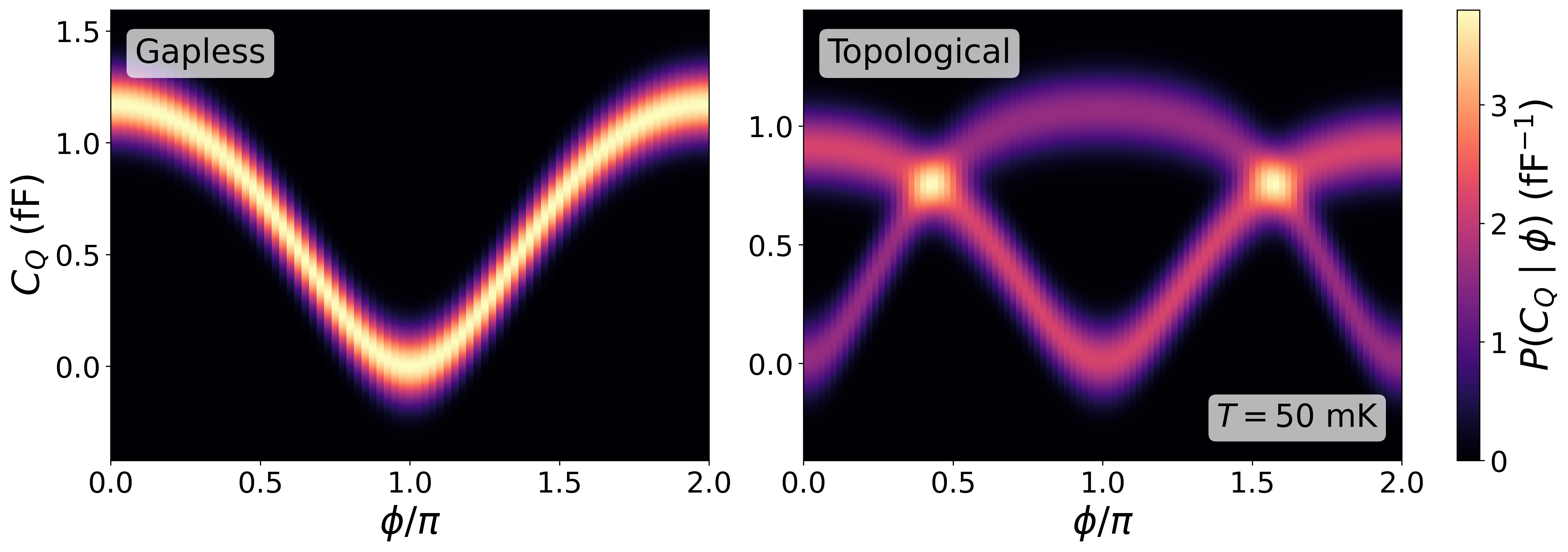}
\caption{\label{fig:clean_boltzmann}
\textbf{Thermally weighted capacitance response in the clean systems.}
Conditional probability density $P(C_Q\mid\phi)$ for the clean gapless wire (left) and topological wire (right) at $T=50~\mathrm{mK}$. The energy cost of occupying the second gapless-wire level strongly suppresses one parity branch, whereas both topological branches retain appreciable weight.}
\end{figure}

\begin{figure*}[t]
\centering
\includegraphics[width=\textwidth]
{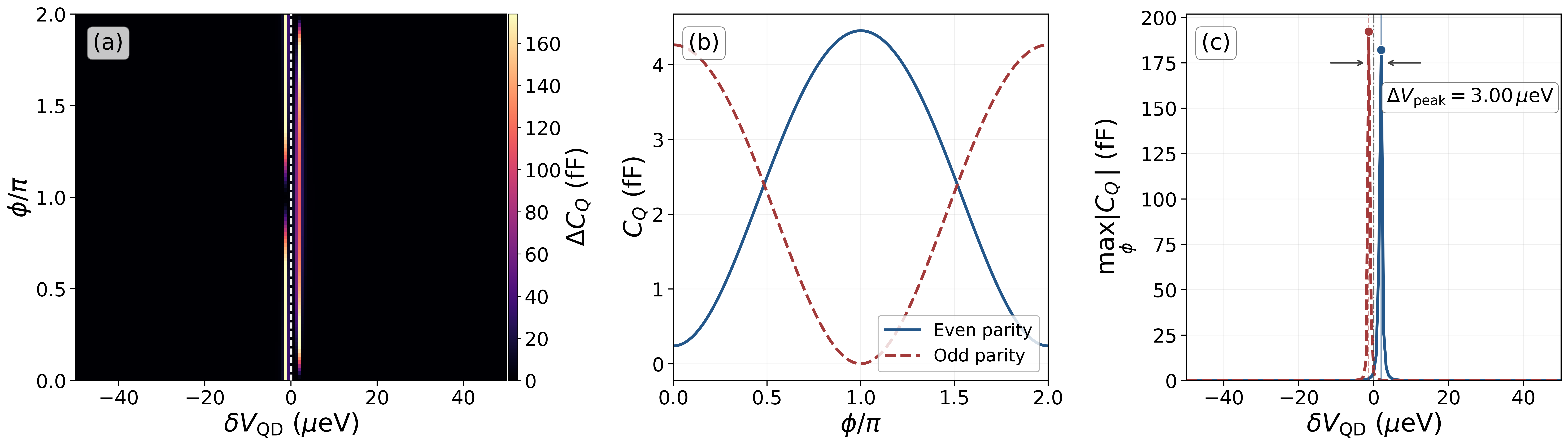}
\caption{\label{fig:normal_fine_tuned}
\textbf{Fine-tuned Majorana-like response of a clean gapless wire.}
(a) Parity contrast $\Delta C_Q$ as a function of $\dVQD$ and $\phi$ near the optimized point. (b) Even- and odd-parity capacitance at $\dVQD=0$. The two $h/e$-periodic branches have comparable amplitudes and are displaced by approximately half a period, producing an apparently $h/2e$-periodic parity contrast. (c) Flux-maximized parity-resolved capacitance. The two resonance maxima are separated by only $3.00\ueV$, but remain distinct, confining equal magnitude of capacitance branches to a narrow gate interval.}
\end{figure*}

For the gapless wire segment, $E_{\mathrm{odd}}=E_\tau=-8.9~\ueV$ and $E_{\mathrm{even}}=E_\tau+E_g=15.54~\ueV$, giving an energy difference of $E_g=24.44\ueV$ between the two sectors. At $T=50~\mathrm{mK}$, the corresponding probabilities are $p_{\mathrm{odd}}^{\mathrm G}=0.9966$ and $p_{\mathrm{even}}^{\mathrm G}=0.0034$. Thus, although the fixed-sector calculation produces a Majorana-like two-branch capacitance response, the energy difference between the two sectors overwhelmingly favors the odd-parity branch. For the topological reference, the unoccupied configuration has $E_{\mathrm{even}}=0$, while the occupied configuration has $E_{\mathrm{odd}}=E_M=1.34\ueV$. The corresponding probabilities are $p_{\mathrm{even}}^{\mathrm T}=0.5769$ and $p_{\mathrm{odd}}^{\mathrm T}=0.4231$. Since the Majorana splitting is smaller than $k_{\mathrm B}T$, both topological parity branches retain appreciable statistical weight.

Fig.~\ref{fig:clean_boltzmann} shows the corresponding conditional probability densities after including the Boltzmann weights at $T=50~\rm mK$. For the gapless wire segment, the weight of one parity sector is strongly suppressed, leaving only one capacitance branch clearly visible. By contrast, both topological parity sectors retain comparable Boltzmann weights, and both capacitance branches remain visible. Thus, in the equilibrium regime where repeated parity switching and relaxation allow the system to sample both sectors according to their respective energies, the relative populations provide an additional distinction between the quantum capacitance responses of the gapless and topological wires.

The strong imbalance between the Boltzmann weights of the two parity sectors in the preceding gapless-wire calculation results from the comparatively large energy separation between them. To give the gapless wire a more favorable test, we consider a second clean parameter set in which the two dot potential resonances are brought much closer together by tuning the system parameters. We use $t=9.18\meV$, $\mu=-0.4975\meV$, $\Gamma_0=0.500\meV$, $w_L=w_R=0.010$, and $\VQD^*=0.5015\meV$, with vanishing spin-orbit coupling, pairing, and disorder. As in the preceding gapless-wire calculations, we retain the $N=1$ and $N=2$ sectors to obtain the most favorable realization of the Majorana-like quantum capacitance response. For this parameter set, the dot potential resonance centers of the two parity sectors are separated by only $3.00\ueV$ , below the thermal scale $k_{\mathrm B}T=4.31\ueV$.

Fig.~\ref{fig:normal_fine_tuned}(a) shows that the resulting parity contrast is confined to a narrow interval of $\VQD$. At $\dVQD=0$, the two fixed-parity curves have comparable amplitudes and are displaced by approximately $\pi$ in $\phi$ [Fig.~\ref{fig:normal_fine_tuned}(b)]. Each branch remains $2\pi$ periodic, corresponding to an $h/e$ flux period, while their absolute difference, $\Delta C_Q$, is approximately $\pi$ periodic. The gapless wire can therefore reproduce an apparently $h/2e$-periodic $\Delta C_Q$ at the optimized reference dot potential. Fig.~\ref{fig:normal_fine_tuned}(c), however, shows that the two underlying dot potential resonances can still be resolved at finer $\VQD$ resolution. Reducing their separation below the thermal scale restores comparable thermal occupation of the energy levels associated with the two parity sectors at the optimized QD potential. At the same time, the balanced capacitance response becomes confined to an even narrower interval along the $\VQD$ axis.

\begin{figure}[h]
\centering
\includegraphics[width=0.8\columnwidth]
{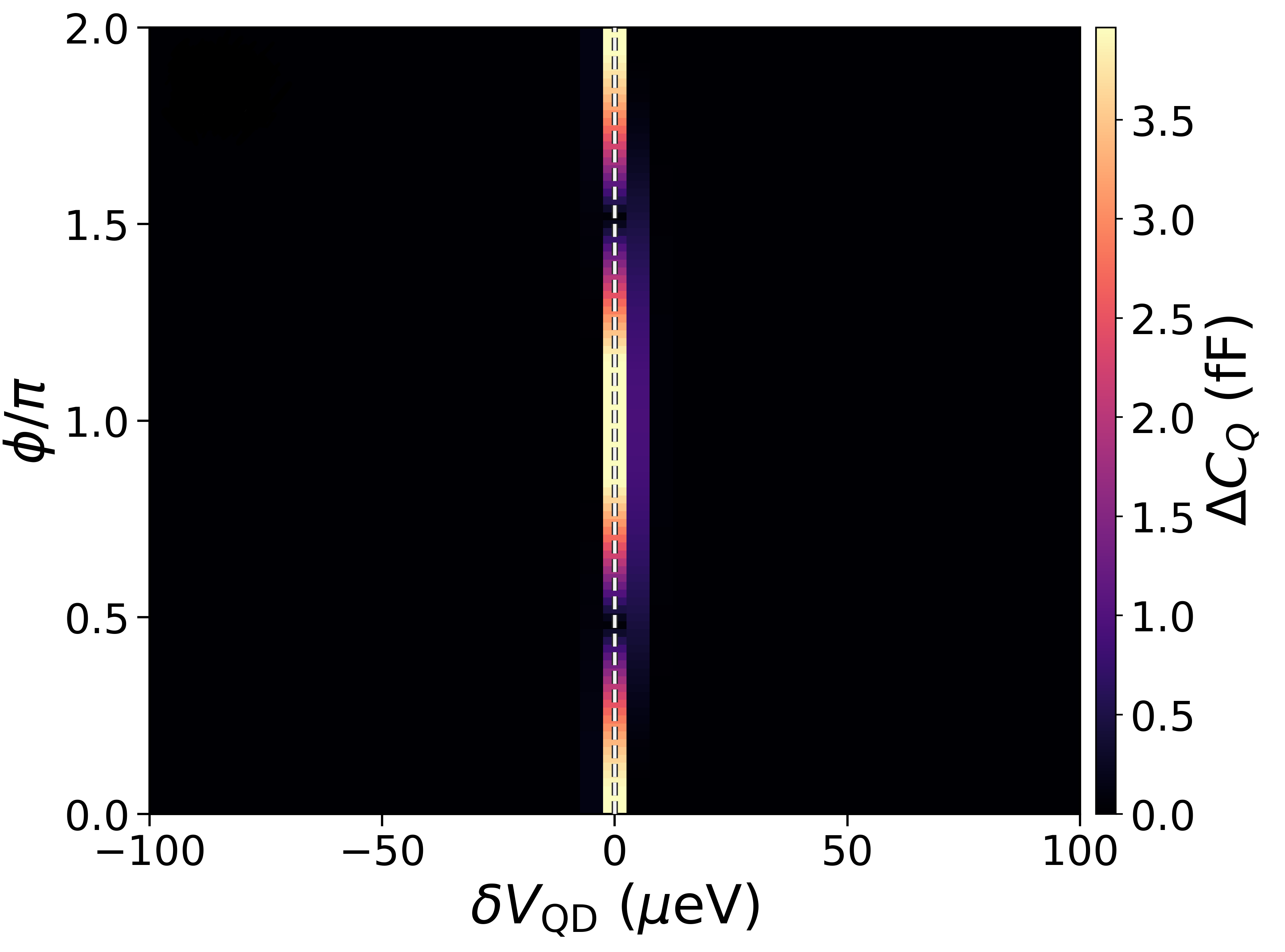}
\caption{\label{fig:finite_resolution}
\textbf{Apparent $h/2e$ periodicity at finite dot potential resolution for the gapless wire as shown in Fig.~\ref{fig:normal_fine_tuned}(a).}
Parity contrast $\Delta C_Q$ for the fine-tuned clean gapless wire after coarse sampling along $\dVQD$. The $3.00\ueV$ separation of the two resonance centers is unresolved, causing the response near the selected gate, marked by the dashed line, to appear as a single $h/2e$-periodic feature.}
\end{figure}

Finite sampling resolution can conceal the narrow separation between the two resonant QD potentials. Fig.~\ref{fig:finite_resolution} shows the same response after coarse sampling along $\dVQD$. At this resolution, the two resonant QD potentials are no longer individually resolved, and the signal near the selected $\VQD$ appears as a single $h/2e$-periodic feature. This apparent periodicity does not make the response robust, since it occurs only within the narrow interval of $\VQD$ where the two parity contributions remain comparable. Even when the individual dot potential resonances cannot be experimentally resolved, a small detuning of $\VQD$ away from the optimized value $\VQD^*$ rapidly unbalances the two parity contributions and destroys the Majorana-like response. Thus, finite gate resolution may obscure the two separate dot resonances, but it does not remove the fine tuning required to produce the Majorana-like response.

The stability of this more strongly tuned point can be tested independently by varying the Zeeman energy without retuning $\VQD$. For this comparison, we calculate the Boltzmann probabilities using the energies of the $N=1$ (odd) and $N=2$ (even) sectors,
\begin{equation}
\begin{aligned}
p_{\mathrm{even}}(\phi;\Gamma)
&=\left[1+e^{\Delta E(\phi;\Gamma)/(k_{\mathrm B}T)}\right]^{-1},\\
\Delta E(\phi;\Gamma)
&=E_{\mathrm{even}}(\phi;\Gamma)
-E_{\mathrm{odd}}(\phi;\Gamma),
\end{aligned}
\label{eq:coupled_boltzmann}
\end{equation}
with $p_{\mathrm{odd}}=1-p_{\mathrm{even}}$. At the optimized Zeeman energy, the flux-averaged probabilities are $\langle p_{\mathrm{even}}\rangle_\phi=51.9\%$ and $\langle p_{\mathrm{odd}}\rangle_\phi=48.1\%$. Changing the Zeeman energy by only $15\ueV$, corresponding to approximately $3\%$ of $\Gamma_0=0.500~\mathrm{meV}$, strongly redistributes the probabilities between the two sectors. For $\delta\Gamma=-15\ueV$, the flux-averaged probabilities become $\langle p_{\mathrm{even}}\rangle_\phi=3.2\%$ and $\langle p_{\mathrm{odd}}\rangle_\phi=96.8\%$, while $\delta\Gamma=+15\ueV$ gives $\langle p_{\mathrm{even}}\rangle_\phi=97.2\%$ and $\langle p_{\mathrm{odd}}\rangle_\phi=2.8\%$. The complete dependence on the Zeeman energy is summarized in Table~\ref{tab:zeeman_boltzmann_weights}.

\begin{figure*}[t]
\centering
\includegraphics[width=\textwidth]
{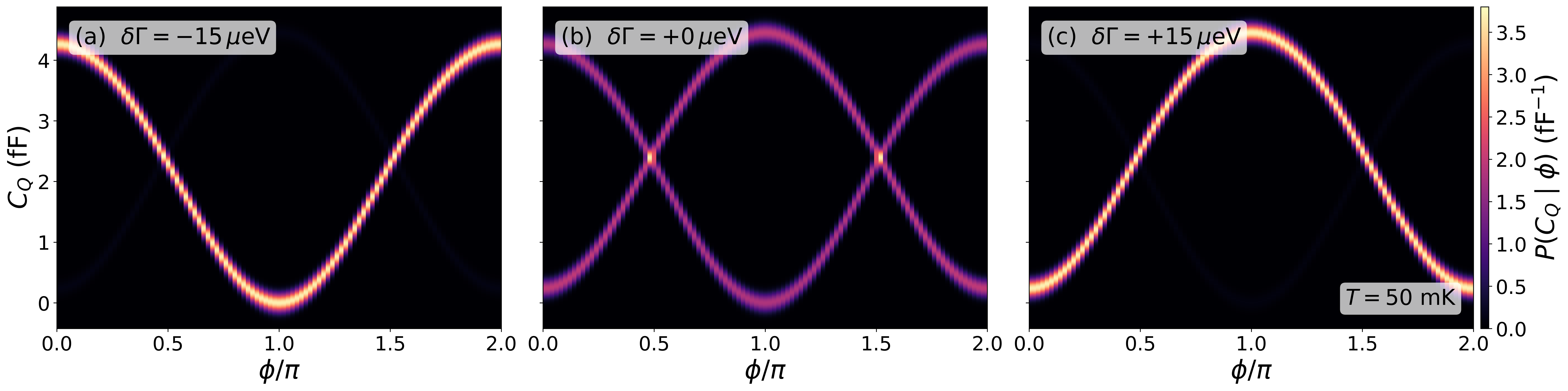}
\caption{\label{fig:zeeman_boltzmann}
\textbf{Zeeman-field sensitivity of the fine-tuned response in gapless wire.}
Conditional capacitance probability density at $T=50~\mathrm{mK}$ for (a) $\delta\Gamma=-15\ueV$, (b) $\delta\Gamma=0$, and (c) $\delta\Gamma=+15\ueV$, where $\delta\Gamma=\Gamma-\Gamma_0$ and $\Gamma_0=0.50\meV$. Both parity sectors have comparable weights at the optimized field, while a $15\ueV$ detuning strongly suppresses one branch. The optimized field gives comparable parity weights, whereas a $15\ueV$ detuning strongly favors one sector and suppresses the other in the thermally weighted response.}
\end{figure*}

Fig.~\ref{fig:zeeman_boltzmann} shows the redistribution of the Boltzmann weights between the parity-resolved capacitance curves of the fine-tuned gapless wire. At $\Gamma=\Gamma_0$, the two parity branches have comparable Boltzmann weights and are simultaneously visible. A detuning of only $\pm15\ueV$, however, strongly favors one parity sector and suppresses the other in the thermally weighted response. The apparently $h/2e$-periodic $\Delta C_Q$ response is therefore restricted not only to a narrow interval of $\VQD$, but also to a narrow interval of $\Gamma$. Together with the chemical-potential dependence established in Sec.~\ref{sec:clean}, this result shows that the simultaneous visibility of the two Boltzmann-weighted parity branches in the gapless wire capacitance response requires fine tuning within the multidimensional $(\VQD,\mu,\Gamma)$ parameter space.

The clean system calculations therefore lead to a consistent conclusion. The original gapless wire reproduces the Majorana-like response only when the QD potential is properly tuned between two distinct narrow dot potential resonances, together with the required tuning of the other parameters that control the separation between the $N=1$ and $N=2$ energy levels. Even when an $h/2e$-periodic $\Delta C_Q$ response is obtained, small detunings of $\VQD$, $\Gamma$, or $\mu$ can produce a strong imbalance between the thermal occupations of the two parity sectors. Bringing the dot potential resonances within the thermal scale restores comparable parity-sector occupations at an optimized point in the $(\VQD,\Gamma,\mu)$ parameter space, but the response remains sharply restricted around that point.

\noindent Appendix~\ref{app:charge_noise} shows that quasistatic fluctuations of the dot potential provide an additional source of instability. These fluctuations disrupt both the comparable amplitudes of the two parity-resolved capacitance responses in the tails of the two QD potential resonance profiles and their approximate half-period shift. Appendix~\ref{app:number_sectors} shows that allowing additional particle number sectors to compete imposes a further restriction. Finally, Appendix~\ref{app:disorder} shows that the distinction between the separated QD resonances of the gapless wire and the common resonance envelope of the topological wire persists in the representative calculation including Rashba spin-orbit coupling and correlated disorder.

\section{Discussion}
\label{sec:conclusion}

We investigate the quantum capacitance of a gapless ($\Delta=0$) normal-wire segment proposed in Ref.~\cite{PinchenkovaKozinHunenbergerLossKlinovaja2026} and find that its Majorana-like response can be experimentally distinguished from the similar capacitance response of a topological Majorana wire. We find that parity-resolved, $h/e$-periodic capacitance signal, with an approximate half-period ($h/2e$) shift between the two parity branches, exists for the normal wire over a narrow range of the quantum-dot potential $\VQD$. This is not, by itself, sufficient to establish nontrivial topology. In the gapless normal wire, the two capacitance branches arise from the resonant coupling of the QD to two different energy levels in the wire. Obtaining comparable magnitudes for the two branches therefore requires the QD energy level to be tuned between the corresponding resonances. This restricts the values of $\VQD$ for which the normal wire displays equal-magnitude capacitance in the two parity sectors, analogous to the topological response, to a narrow region along the $\VQD$ axis. 

\begin{table}[h]
\centering
\caption{Flux-averaged parity weights at $T=50~\mathrm{mK}$ for the fine-tuned parameter space point in the clean gapless wire. Small detunings from $\Gamma_0=0.500~\mathrm{meV}$ strongly favor one parity sector.}
\label{tab:zeeman_boltzmann_weights}
\renewcommand{\arraystretch}{1.15}
\setlength{\tabcolsep}{3pt}
\begin{tabular}{c c c c}
\toprule
$\delta\Gamma$ & $\Gamma$
& $\langle p_{\mathrm{odd}}\rangle_\phi$
& $\langle p_{\mathrm{even}}\rangle_\phi$\\
($\mu\mathrm{eV}$) & ($\mathrm{meV}$) & (\%) & (\%)\\
\midrule
$-50$ & $0.450$ & $99.9990$ & $0.0010$\\
$-30$ & $0.470$ & $99.8978$ & $0.1022$\\
$-15$ & $0.485$ & $96.7831$ & $3.2169$\\
$0$   & $0.500$ & $48.0690$ & $51.9310$\\
$+15$ & $0.515$ & $2.7689$  & $97.2311$\\
$+30$ & $0.530$ & $0.0875$  & $99.9125$\\
$+50$ & $0.550$ & $0.0008$  & $99.9992$\\
\bottomrule
\end{tabular}
\end{table}

\noindent This is fundamentally different from the topological superconducting case, where the two capacitance branches correspond to the two occupations of the same low-energy fermionic mode constructed from the end Majoranas and are therefore centered at nearly identical values of the QD potential. Consequently, the values of $\VQD$ for which bimodal, equal-amplitude capacitance oscillations, with $\Delta C_Q$ exhibiting $h/2e$ periodicity, can be observed extend over a broad range. Thus, although the two systems can generate similar flux-dependent signals at a suitably chosen value of $\VQD$, they respond very differently as $\VQD$ and other experimentally controlled parameters are varied. This stark difference in parameter dependence, where Majorana-like capacitance oscillations occur over a broad range of $\VQD$ in the topological wire but only at a fine-tuned value of the $\VQD$ in the normal-wire segment, provides a direct experimental distinction between the two cases.

For the clean wire results, this difference can be seen directly from the dependence of the capacitance response on $\VQD$. As shown in the top panel of Fig.~\ref{fig:clean_maps}, the two dot potential resonances in the gapless wire, corresponding to two parity sectors, are separated by $\sim35.0\ueV$, which is close to the $\sim33.3\ueV$ separation between the two relevant energy levels in the normal wire segment, $E_\tau$ and $E_g$. In the topological case, the corresponding dot potential resonance separation is only $\sim5.25\ueV$ and is set by the energy scale $4E_M$, where $E_M$ is the Majorana splitting energy, as shown in the bottom panel of Fig.~\ref{fig:clean_maps}. Since $E_M$ is generally small for robust MZMs, the two parity-resolved resonance centers lie close together, leading to both capacitances appearing as equal intensity oscillations over a broad range of dot potential $\VQD$. This means, in the topological case, there can be multiple dot potentials that support Majorana-like capacitance oscillations. This comparison can already be performed experimentally. In Ref.~\cite{MicrosoftAzureQuantumNature2025}, the flux-dependent capacitance was measured as a function of the QD plunger voltage. The measurements show bimodal quantum capacitance oscillations across several successive dot transitions, with the visibility varying between transitions [see Fig.~3(i) and Supplementary Fig.~S11 of Ref.~\cite{MicrosoftAzureQuantumNature2025}]. This parameter-space stability of the Majorana-like oscillations of the quantum capacitance in the experiments helps rule out a gapless normal wire segment as responsible for the oscillations. 

The magnitude of the capacitance difference provides an additional comparison between the normal and topological wire. As shown in Fig.~\ref{fig:clean_peaks} left panel, at the fine-tuned dot potential ($\dVQD=0$) where the two parity branches of capacitance for the normal wire have comparable amplitudes, $\max_\phi\Delta C_Q(\dVQD=0)$ is of order $\sim1~\mathrm{fF}$. Moving $\VQD$ toward either resonance can increase $\Delta C_Q$, but one parity branch then rapidly becomes dominant and the Majorana-like response is lost. In contrast, for the topological wire the magnitude of the difference between the parity branches of the capacitance response reaches approximately $\sim10~\mathrm{fF}$ near its common resonance, as seen in Fig.~\ref{fig:clean_peaks} right panel. According to Fig.~\ref{fig:realistic_peaks} in Appendix~\ref{app:disorder} the difference in magnitude of $\Delta C_Q$ between the normal and topological superconducting wires, for the relevant dot potentials, is even more pronounced in the presence of disorder.

%The relevant comparison is therefore between the largest parity contrast compatible with a Majorana-like response in the gapless wire and the parity contrast obtained near the approximately common resonant value of $\VQD$ in the topological wire. 

%Although these absolute values depend on the system parameters, the gapless wire response is subject to a more general tradeoff. Its Majorana-like flux dependence appears only between the two resonant values of $\VQD$, where the two parity branches have comparable amplitudes. 

Varying the system parameters other than the QD potential $\VQD$ makes the fine tuning required to obtain the Majorana-like quantum capacitance response in the gapless normal wire segment more apparent. As shown in Fig.~\ref{fig:mu_detuning}, a change in $\mu$ shifts both resonant QD-potential values associated with the two parity sectors relative to a fixed reference QD potential, requiring a corresponding adjustment of the reference dot potential $\VQD^*$ (see Eq.~\ref{eq:VQD-detuning}) to keep their contributions comparable. By contrast, in the topological wire, the quantum capacitance response remains centered within nearly the same gate region over the entire range $-50\ueV\leq\delta\mu\leq50\ueV$. Therefore to maintain a Majorana-like response in the normal wire, a change in chemical potential $\mu$ by the plunger gate would necessitate a corresponding change of $\VQD^*$, whereas in the topological superconducting wire, the Majorana-like response of quantum capacitance is much more robust to changes in $\mu$. 

The two systems also behave differently at finite temperature. Since $E_\tau$ and $E_g$ energy levels in the normal wire segment are separated by $\sim 33~\ueV$, which is much larger than the thermal energy $k_B T\sim4.31~\ueV$ at $T=50~\mathrm{mK}$, in the equilibrium limit, more than $99\%$ of the total Boltzmann weight (and hence the intensity) lies in only one of the two relevant parity sectors. In contrast, at $50~\rm mK$, since the two parity sectors are nearly degenerate in energy the capacitance of both odd and even parity retain appreciable weight in the topological case [see Fig.~\ref{fig:clean_boltzmann} in Sec.~\ref{sec:clean_visibility}]. 

To determine whether the imbalance in intensity between the two capacitance branches in the normal wire can be avoided by reducing the separation between the relevant energy levels for $N=1$ (odd parity sector) and $N=2$ (even parity sector) particle number sectors, we also consider a second parameter set for the gapless normal wire segment in which the resonant QD-potential values associated with the two parity sectors are separated by only $3.00\ueV$, smaller than the thermal energy. At the optimized dot potential, the two parity sectors now have comparable intensities, and the resulting parity contrast $\Delta C_Q$ in capacitance can appear $h/2e$ periodic when the two resonant QD-potential values are not separately resolved at finite dot potential resolution [see Figs.~\ref{fig:normal_fine_tuned} and \ref{fig:finite_resolution} in Sec.~\ref{sec:clean_visibility}]. This does not make the quantum capacitance response of the gapless normal wire segment more robust. Instead, the Majorana-like response is restricted to an even narrower range of $\VQD$, and changing the Zeeman energy by only $\sim15\ueV$, approximately $3\%$ of the optimized value $\Gamma_0=0.50~\meV$, transfers most of the statistical weight to one parity sector [see Table~\ref{tab:zeeman_boltzmann_weights} in Sec.~\ref{sec:clean_visibility}]. Maintaining the Majorana-like response in the gapless normal wire segment therefore requires the simultaneous tuning of $\VQD$, $\mu$, and $\Gamma$.

The calculations presented in the Appendices show that the conditions for obtaining the Majorana-like response in the gapless wire become even more restrictive when charge fluctuations, additional particle number sectors, and disorder are included. Quasistatic fluctuations of the dot potential disturb the comparable amplitudes and approximate half-period shift of the two parity-resolved capacitance curves; see Fig.~\ref{fig:charge_noise_deltaC} in Appendix~\ref{app:charge_noise}. In the main text calculations, we restrict the particle number sectors to $N=1$ and $N=2$, which provides the most favorable conditions for obtaining the Majorana-like quantum capacitance response in the gapless wire segment. When additional sectors are included, the $N=0$ or $N=3$ state can become energetically preferred, causing the capacitance branch associated with one of the required parity sectors to disappear, shown in Fig.~\ref{fig:unrestricted_deltaC} in Appendix~\ref{app:number_sectors}. Finally, including disorder does not change the main result, the parity sector resolved resonances of the gapless normal wire remain separated, while those of the topological wire remain within, approximately, the same resonant $\VQD$ range as illustrated in Fig.~\ref{fig:realistic_peaks}, Appendix~\ref{app:disorder}.

These results can also be understood from the interferometric perspective discussed in the Introduction. The capacitance response probes two defining properties of a Majorana state: a low-energy fermionic mode and its nonlocal spatial structure. A nearly parity-independent QD resonance position, together with comparable occupation of the two capacitance branches in the telegraph signal, provides evidence for a single low-energy mode associated with two different parity sectors. Such a combination does not occur generically in a gapless non-superconducting wire without fine tuning. At the same time, coherent single-electron interference is naturally $h/e$ periodic. In the ideal topological limit, insertion of an $h/2e$ superconducting flux quantum changes the ground-state fermion parity and maps one parity branch onto the other~\cite{PUKitaev2001,PRLFu2010,PRBSauSwingleTewari2015,MicrosoftAzureQuantumNature2025,RoyCasonSharmaTewari2026}. The gapless wire segment of Ref.~\cite{PinchenkovaKozinHunenbergerLossKlinovaja2026} shows that an apparent $h/2e$ relation between two $h/e$-periodic branches can also arise without superconductivity. However, this requires a particular alignment of the energy levels associated with the relevant particle-number sectors and their tunnel amplitudes. Our results show that this relation is correspondingly sensitive to independent variations of the system parameters. A separate ambiguity in the present capacitance measurements is that the two observed branches are inferred, rather than independently verified, to correspond to opposite fermion parities. This leaves room for non-topological scenarios involving narrowly avoided crossings or partially separated low energy states~\cite{StanescuTewariPRB2026,RoySauTewari2026}. Such alternatives, however, require the avoided gap or spatial overlap to remain anomalously small over the relevant tuning range and are therefore less generic than a protected parity crossing associated with a stable topological region. Given the observed persistence of the capacitance response over multiple dot transitions and control parameter settings~\cite{MicrosoftAzureQuantumNature2025,MicrosoftQuantumPb2026,boutin2025predictivesimulationsdynamicalresponse}, the natural interpretation is that the two $h/e$-periodic capacitance branches correspond to opposite fermion-parity sectors. In this case, an $h/2e$ flux insertion that maps one parity branch onto the other is consistent with spatially separated Majorana modes localized near the two ends of the wire. Departures from this ideal $h/2e$ relation are expected when the Majorana modes cease to be well localized at opposite ends and acquire appreciable overlap across the finite wire, as occurs in the finite-size and disordered ``topological patch'' regime~\cite{DasSarmaSauStanescuPRB2023,StanescuTewariPRB2026}.

%Our results demonstrate that the specific two-level approximation for the gapless wire segment is directly falsifiable experimentally. To account for the equal-intensity, bimodal quantum capacitance response, with $\Delta C_Q$ having an approximate period of $h/2e$, the gapless normal wire segment must be fine-tuned to isolated optimized points in the parameter space in particular to a very narrow range in the QD-potential $\VQD$ preserve the simultaneous visibility of both parity branches, their characteristic flux dependences, and an appreciable parity-contrast magnitude as $\VQD$, $\mu$, and $\Gamma$ are varied independently. In the equilibrium limit, the relative visibility of the two branches must also remain consistent with their finite-temperature Boltzmann weights. The response must further remain stable against charge fluctuations and the participation of experimentally accessible particle number sectors. These requirements are substantially stronger than reproducing an optimized flux trace at a single fine-tuned point in the parameter space for a gapless normal wire segment.

\section{Conclusion}
We conclude that, in a gapless normal-wire segment, bimodal, equal-intensity oscillations of the quantum capacitance, with $\Delta C_Q$ exhibiting an approximate $h/2e$ periodicity, occur only within a narrow interval of the QD potential $\VQD$. This narrow range results from the finite energy separation, approximately $33.3~\ueV$ for the parameters used both here and in Ref.~\cite{PinchenkovaKozinHunenbergerLossKlinovaja2026}, between the even- and odd-parity resonances as functions of $\VQD$. In contrast, the analogous Majorana-like capacitance response of the topological wire persists over a broad range of parameters, including a broad range of $\VQD$. Furthermore, in the clean case shown in Fig.~2, the normal-wire capacitance signal is approximately an order of magnitude weaker than its topological counterpart at the value of $\VQD$ where the even- and odd-parity signals have equal magnitude. The discrepancy becomes even larger in the presence of disorder, as shown in Fig.~\ref{fig:realistic_peaks} of the Appendix~\ref{app:disorder}, providing an additional quantitative distinction between the two scenarios. Recent experiments~\cite{MicrosoftAzureQuantumNature2025,MicrosoftQuantumPb2026} observe $h/2e$-periodic oscillations of $\Delta C_Q$ across many quantum-dot transitions and at several values of $\VQD$. We therefore conclude that this robustness in parameter space rules out a gapless normal-wire segment as the origin of the observed Majorana-like capacitance response.

\begin{acknowledgments}
S.T. and B.B.R acknowledge support from ARO Grant No. ARO-W911NF2210247. J.S. acknowledges support from the Joint Quantum Institute and the Laboratory for Physical Sciences through the Condensed Matter Theory Center.
\end{acknowledgments}

\appendix

\begin{figure}[h]
\centering
\includegraphics[width=\linewidth]{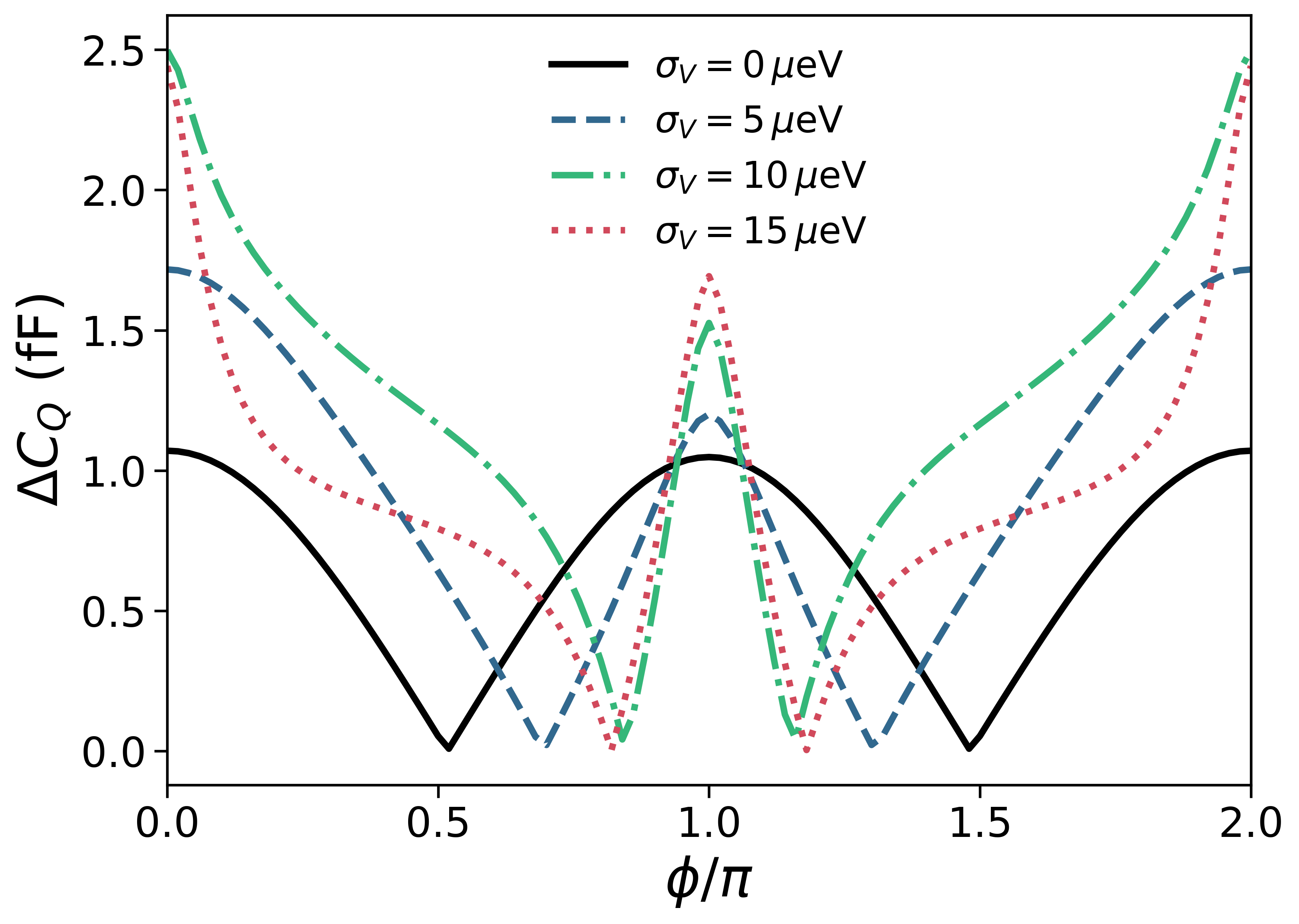}
\caption{\label{fig:charge_noise_deltaC}
\textbf{Charge-noise sensitivity of the clean gapless-wire response.}
Parity contrast obtained after separately averaging the even- and odd-parity capacitances over quasistatic Gaussian fluctuations of the dot potential. The mean is fixed at $\VQD^*=0.585\meV$, and $\sigma_V$ denotes the noise amplitude in energy units. The clean response is approximately $h/2e$ periodic. Increasing $\sigma_V$ makes the responses near $\phi=0$, $\pi$, and $2\pi$ unequal and shifts the capacitance nodes, disrupting the fine-tuned periodic relation without necessarily suppressing the overall signal. All other parameters are those of the clean gapless wire.}
\end{figure}

\section{Sensitivity of the gapless-wire response to charge noise}
\label{app:charge_noise}

The Majorana-like response of the clean gapless wire requires the dot level to lie between the two relevant wire resonances, where the even- and odd-parity capacitance contributions have comparable amplitudes. We test the stability of this condition against quasistatic fluctuations of the dot potential. Taking the mean operating point to be $\VQD^*=0.585\meV$, we define the noise-averaged response in parity sector $p$ as
\begin{equation}
\overline{C}_Q^{(p)}(\phi;\sigma_V)
=
\int dV\,
\frac{e^{-(V-\VQD^*)^2/(2\sigma_V^2)}}
{\sqrt{2\pi}\sigma_V}
C_Q^{(p)}(\phi,V),
\label{eq:charge_noise_average}
\end{equation}
and the corresponding parity contrast as
\begin{equation}
\Delta C_Q(\phi;\sigma_V)
=
\left|
\overline{C}_Q^{(\mathrm{even})}(\phi;\sigma_V)
-
\overline{C}_Q^{(\mathrm{odd})}(\phi;\sigma_V)
\right|.
\label{eq:charge_noise_deltaC}
\end{equation}
The averaging is performed separately in the two parity sectors before their difference is evaluated. Here $\sigma_V$ characterizes quasistatic fluctuations of the dot potential and is distinct from the capacitance-readout broadening $\sigma_C$ introduced in Sec.~\ref{sec:clean_visibility}.

Fig.~\ref{fig:charge_noise_deltaC} shows the resulting flux dependence for $\sigma_V=0$, $5$, $10$, and $15\ueV$, with $\sigma_V=0$ denoting the unaveraged result. The largest fluctuation scale is comparable to the detunings from the operating gate to the two parity-resolved capacitance maxima, approximately $14\ueV$ and $19\ueV$. In the absence of charge noise, $\Delta C_Q$ approximately repeats under $\phi\rightarrow\phi+\pi$, producing the apparent $h/2e$-periodic response. As $\sigma_V$ increases, this relation is progressively lost: the responses near $\phi=0$, $\pi$, and $2\pi$ develop unequal amplitudes, the nodes shift from their clean-system positions, and an additional narrow feature develops near $\phi=\pi$.

The noise average does not necessarily reduce $\Delta C_Q$ monotonically, because the fluctuating dot potential can sample values closer to one of the two narrow parity resonances, where the capacitance is larger. The relevant effect is instead the loss of the balanced relation between the two parity contributions. Quasistatic charge fluctuations therefore provide an additional constraint on the gapless wire construction: rather than stabilizing the apparent $h/2e$-periodic response, they expose its sensitivity to the electrostatic operating point.

\begin{figure}[t]
\centering
\includegraphics[width=0.90\linewidth]
{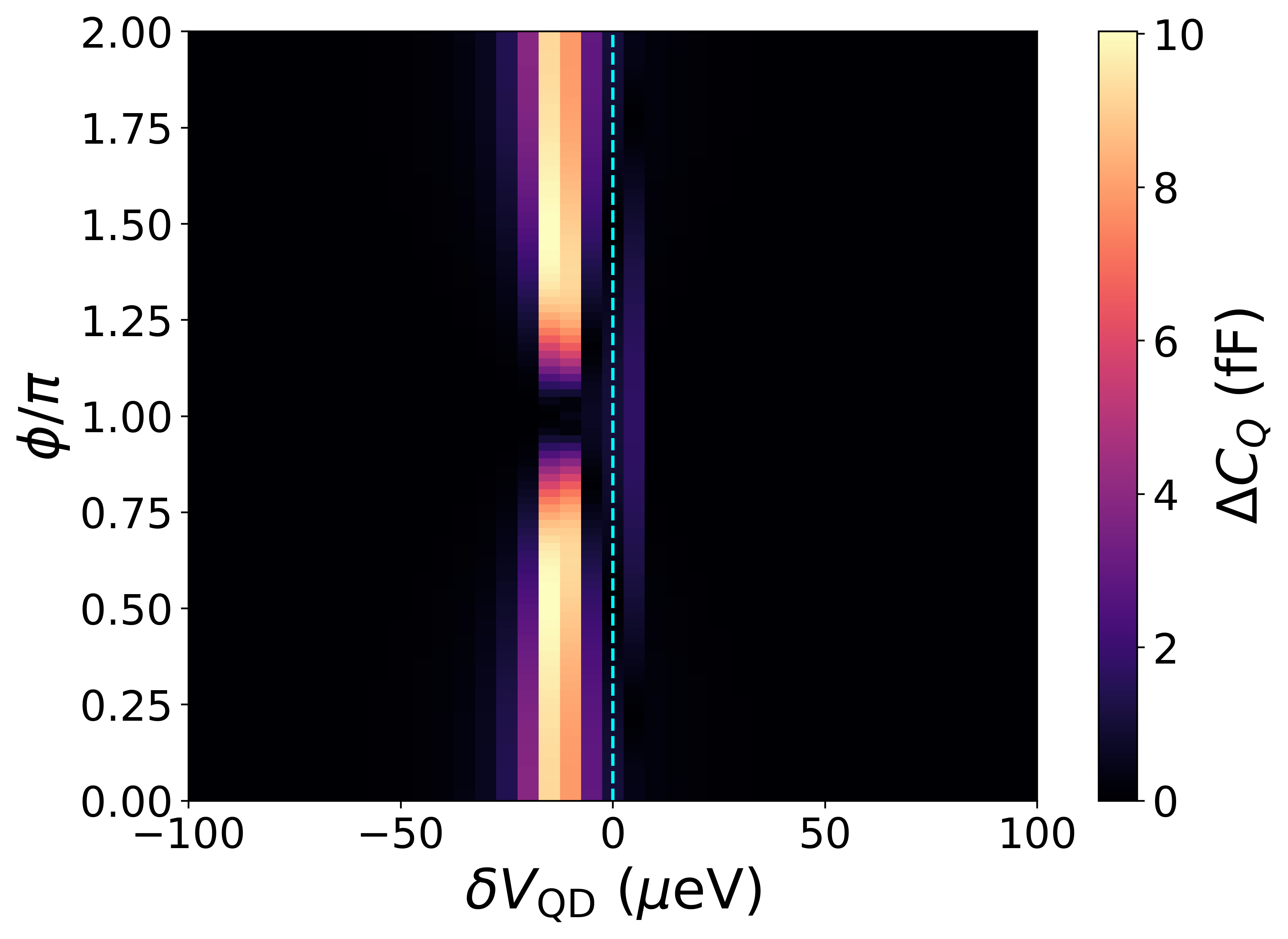}
\caption{\label{fig:unrestricted_deltaC}
\textbf{Capacitance response with unrestricted particle number sectors for the clean gapless wire.} 
Parity contrast of the clean gapless wire when the even- and odd-parity branches are selected as the lowest-energy states over all even and odd particle numbers, respectively. The horizontal axis is the detuning from $\VQD^*=0.585\meV$, and the cyan dashed line marks the operating gate used for the fixed-$N=1,2$ calculation. Allowing unrestricted particle numbers suppresses the even parity resonance observed earlier at $\delta V_{QD}=+19.4\,\ueV$, demonstrating the dependence of the Majorana-like pattern on the fixed-sector prescription.}
\end{figure}

\section{Dependence on the accessible particle number sectors}
\label{app:number_sectors}

The clean gapless normal wire comparison in the main text follows the fixed particle number sector prescription, in which the odd- and even-parity branches are identified with the $N=1$ and $N=2$ sectors, respectively. This choice isolates the two nontrivial particle-number sectors responsible for the Majorana-like response. Within the corresponding three-level low-energy model, the empty ($N=0$) and fully occupied ($N=3$) sectors have vanishing quantum capacitance~\cite{PinchenkovaKozinHunenbergerLossKlinovaja2026}. An electrostatic charging contribution of the form $E_C(N-n_g)^2$ can, in principle, energetically favor a desired pair of adjacent charge sectors. Such a term is not included explicitly here. The fixed-$N=1,2$ prescription should therefore be understood as a favorable restriction that prevents the $N=0$ and $N=3$ sectors from suppressing the Majorana-like quantum capacitance response of the gapless-wire model.

\begin{table}[t]
\centering
\caption{Flux-averaged probabilities of the $N=0,\ldots,3$ particle number sectors as functions of dot potential detuning for the clean gapless wire at $T=50~\mathrm{mK}$. The chemical potential is fixed at $\mu=-0.480\meV$, with $\VQD=0.585\meV+\dVQD$.}
\label{tab:VQD_sector_weights}
\setlength{\tabcolsep}{7pt}
\renewcommand{\arraystretch}{1.15}
\begin{tabular}{rcccc}
\toprule
$\dVQD$
& $\langle P_0\rangle_{\phi}$
& $\langle P_1\rangle_{\phi}$
& $\langle P_2\rangle_{\phi}$
& $\langle P_3\rangle_{\phi}$\\
($\mu\mathrm{eV}$)
& (\%) & (\%) & (\%) & (\%) \\
\midrule
$-50$ & 0.0003  & 11.6224 & 88.0972 & 0.2801  \\
$-25$ & 0.0890  & 12.9377 & 86.7087 & 0.2647 \\
$-15$ & 0.7744  & 18.8744 & 80.1144 & 0.2368 \\
$-5$  & 4.3687  & 45.9489 & 49.5446 & 0.1378 \\
$0$   & 7.1192  & 65.4005 & 27.4082 & 0.0722 \\
$+5$  & 9.0848  & 78.6209 & 12.2645 & 0.0297 \\
$+15$ & 10.5239 & 87.2707 & 2.2018  & 0.0036 \\
$+25$ & 10.8297 & 88.4742 & 0.6957  & 0.0004 \\
$+50$ & 11.0097 & 88.6219 & 0.3683  & 0.0000 \\
\bottomrule
\end{tabular}
\end{table}

\begin{table}[h]
\centering
\caption{Flux-averaged probabilities of the $N=0,\ldots,3$ particle number sectors at $T=50~\mathrm{mK}$ under compensated chemical-potential detuning. The parameters are varied as $\mu=\mu_0+\delta\mu$ and $\VQD=\VQD^*-\delta\mu$, with $\mu_0=-0.480\meV$ and $\VQD^*=0.585\meV$.}
\label{tab:compensated_mu_sector_weights}
\renewcommand{\arraystretch}{1.15}
\setlength{\tabcolsep}{4pt}
\begin{tabular}{c c c c c c}
\toprule
$\delta\mu$
& $\VQD$
& $\langle P_0\rangle_{\phi}$
& $\langle P_1\rangle_{\phi}$
& $\langle P_2\rangle_{\phi}$
& $\langle P_3\rangle_{\phi}$\\
($\mu\mathrm{eV}$)
& ($\mathrm{meV}$)
& (\%) & (\%) & (\%) & (\%)\\
\midrule
$-50$ & $0.635$ & $99.9916$ & $0.0084$  & $0.0000$ & $0.0000$\\
$-25$ & $0.610$ & $97.2957$ & $2.7009$  & $0.0034$ & $0.0000$\\
$-15$ & $0.600$ & $77.7485$ & $21.9680$ & $0.2835$ & $0.0000$\\
$-5$  & $0.590$ & $23.5066$ & $67.5790$ & $8.9071$ & $0.0073$\\
$0$   & $0.585$ & $7.1192$  & $65.4005$ & $27.4082$ & $0.0722$\\
$+5$  & $0.580$ & $1.4339$  & $42.1494$ & $55.9425$ & $0.4742$\\
$+15$ & $0.570$ & $0.0214$  & $6.4210$  & $86.0630$ & $7.4947$\\
$+25$ & $0.560$ & $0.0001$  & $0.3848$  & $53.0411$ & $46.5740$\\
$+50$ & $0.535$ & $0.0000$  & $0.0000$  & $0.3480$ & $99.5713$\\
\bottomrule
\end{tabular}
\end{table}

We first remove this restriction and select the lowest-energy state independently within each parity sector. For ordered single-particle energies $\epsilon_j$, the minimum energy at fixed particle number is
\begin{equation}
E_N(\phi,\VQD)=\sum_{j=1}^{N}\epsilon_j(\phi,\VQD),
\label{eq:particle_number_energy}
\end{equation}
and the particle numbers associated with the two parity ground states are
\begin{equation}
N_{\rm even}=\underset{N\ {\rm even}}{\operatorname{argmin}}\,E_N,
\qquad
N_{\rm odd}=\underset{N\ {\rm odd}}{\operatorname{argmin}}\,E_N.
\label{eq:unrestricted_particle_numbers}
\end{equation}
For comparison with the restricted particle-number-sector result shown in Fig.~\ref{fig:clean_maps}, Fig.~\ref{fig:unrestricted_deltaC} shows the parity contrast of the gapless wire when all particle-number sectors are allowed to compete. Allowing the particle number within each parity sector to be determined by energy minimization substantially changes the quantum capacitance response in the $(\phi,\VQD)$ plane. In particular, over part of the $\dVQD$ range associated with the fixed-$N=2$ resonance in the even-parity sector, the lowest-energy even-parity state instead belongs to the $N=0$ sector. Its quantum capacitance vanishes within the low-energy model, removing the even-parity resonance at $\dVQD=+19.4\,\ueV$. The strongest remaining contrast then comes from the odd-parity response near $\dVQD=-13.9\,\ueV$ and is suppressed near $\phi=\pi$. Thus, neither the optimized reference dot potential nor the detailed flux profile of the fixed-$N=1,2$ construction is necessarily retained when the energetically preferred particle-number sectors are allowed to vary.

\begin{figure}[t]
\centering
\includegraphics[width=0.88\columnwidth]{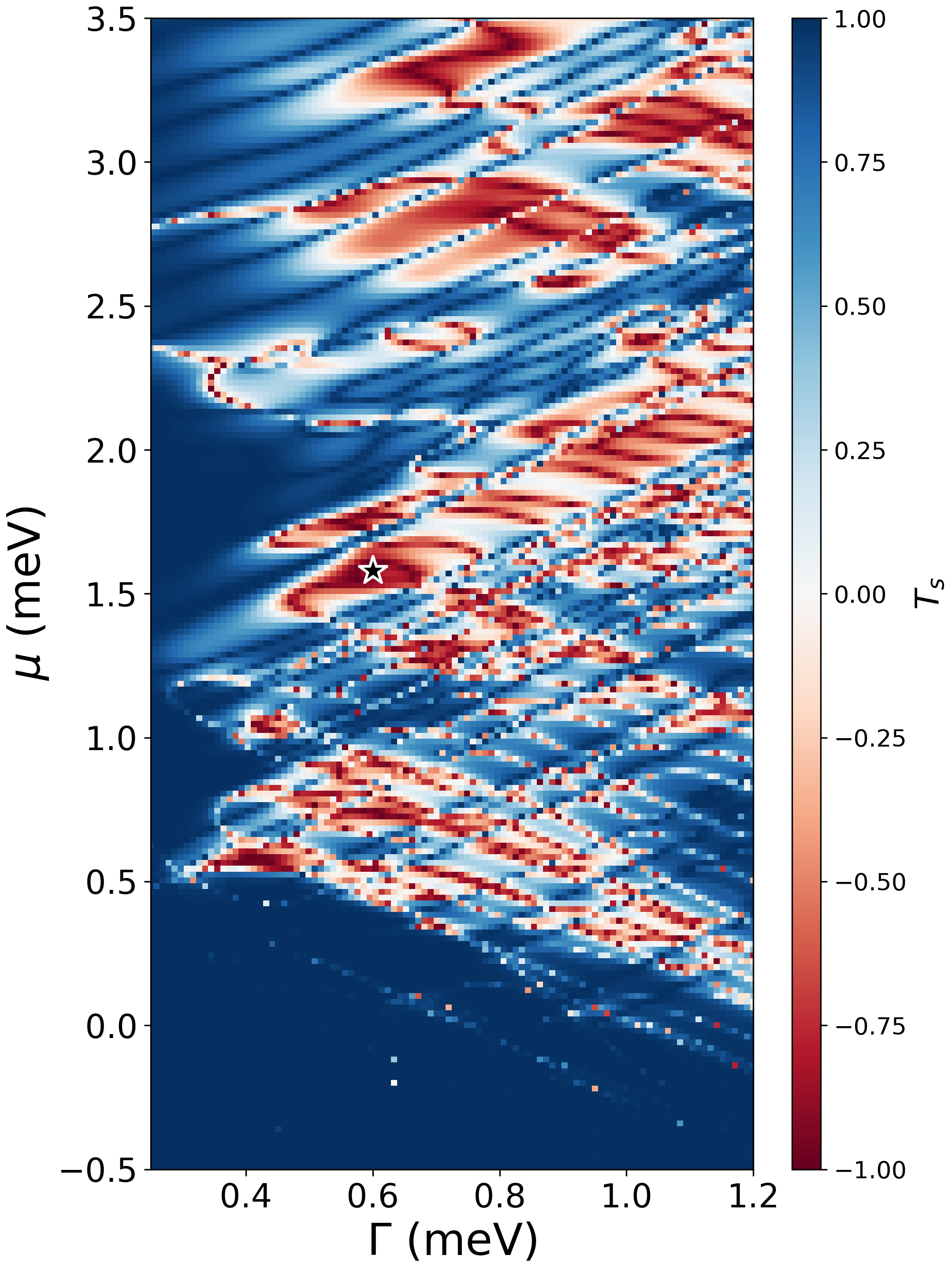}
\caption{\label{fig:Ts_map}
\textbf{Topological stability map for the disordered superconducting wire.}
The signed quantity $T_s$ is shown as a function of $\Gamma$ and $\mu$ for $\alpha=1.4\meV$, $V_0=1.2\meV$, and $\xi_{\rm dis}=20~\mathrm{nm}$. Blue and red denote the trivial ($T_s>0$) and topological ($T_s<0$) sectors, respectively, while $|T_s|\simeq0$ identifies regions with strong boundary condition sensitivity. The star marks the operating point $(\Gamma,\mu)=(0.60,1.58)\meV$.}
\end{figure}

The preceding calculation uses only the lowest-energy state in each parity sector and therefore corresponds to the zero-temperature limit. At finite temperature, higher-energy states and states with different particle numbers can also be occupied. To calculate the probability of occupying a sector with particle number $N$, we first sum the Boltzmann weights of all many-body states in that sector,
\begin{equation}
\begin{aligned}
Z_N(\phi,\VQD)
&=\sum_{\alpha\in\mathcal H_N}
e^{-E_{N\alpha}(\phi,\VQD)/(k_{\mathrm B}T)},\\
P_N&=\frac{Z_N}{\sum_M Z_M},
\end{aligned}
\label{eq:sector_probability}
\end{equation}
where $\alpha$ labels the many-body states belonging to the $N$-particle sector. The total probabilities of occupying an even- or odd-parity sector are then
\begin{equation}
P_{\rm even}=\sum_{N\ {\rm even}}P_N,
\qquad
P_{\rm odd}=\sum_{N\ {\rm odd}}P_N.
\label{eq:total_parity_probabilities}
\end{equation}
The probabilities reported below are calculated at $T=50~\mathrm{mK}$ and averaged over flux. This calculation assumes that the system can sample states with different particle numbers during the measurement. If the total charge is fixed, only the corresponding particle-number sector contributes. These probabilities should not be confused with the two number sector Boltzmann weights used in Sec.~\ref{sec:clean_visibility}. In that calculation, only the restricted particle number sectors $N=1$ and $N=2$ sectors were included in order to reproduce the original gapless wire bench-marked calculations under the most favorable conditions.

\begin{figure*}[t]
\centering
\includegraphics[width=0.93\textwidth]{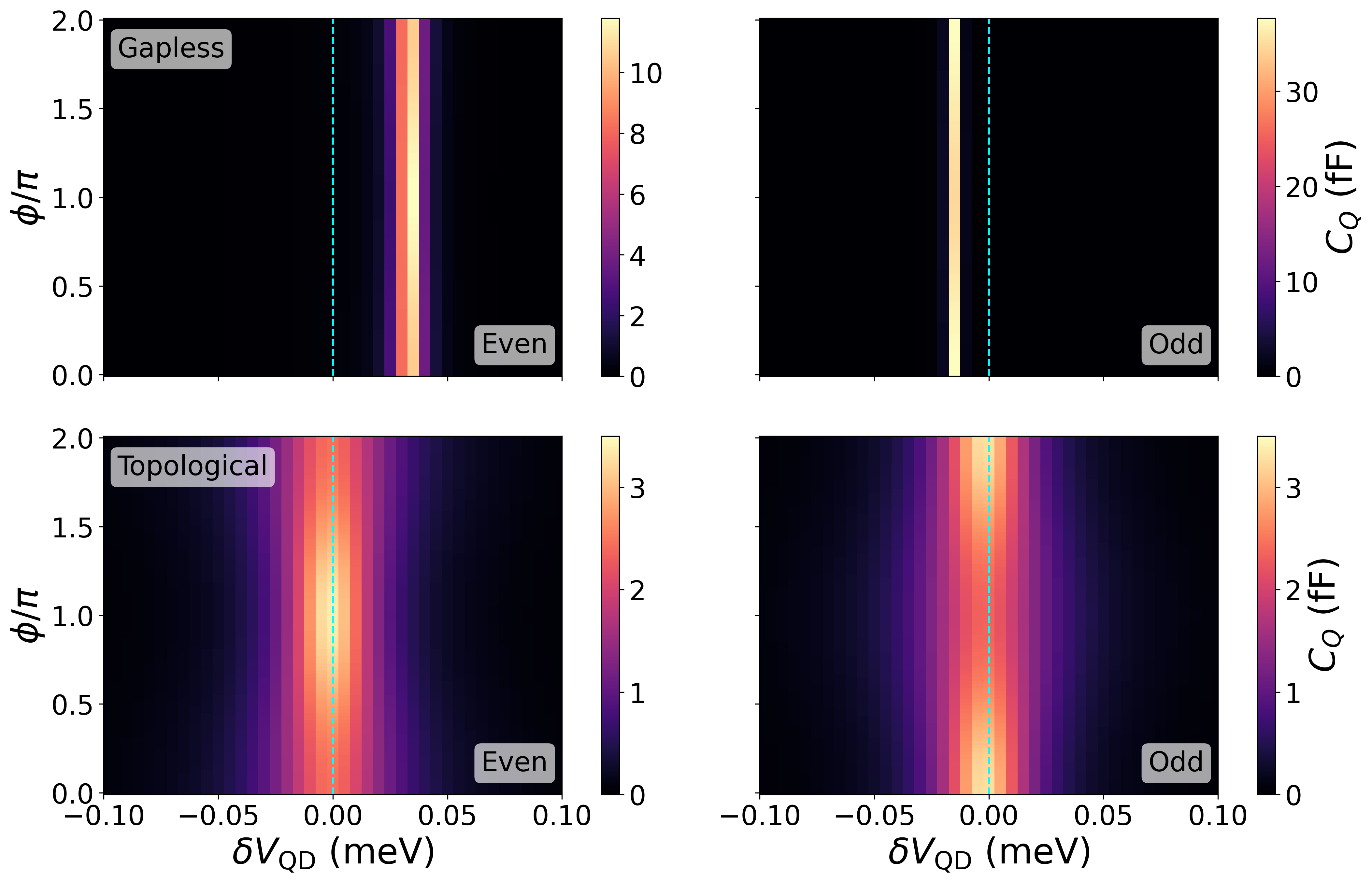}
\caption{\label{fig:realistic_maps}
\textbf{Parity-resolved capacitance with Rashba coupling and disorder.}
Absolute quantum capacitance as a function of $\dVQD$ and $\phi$ at $(\Gamma,\mu)=(0.60,1.58)\meV$, with $\alpha=1.4\meV$, $V_0=1.2\meV$, and $\xi_{\rm dis}=20~\mathrm{nm}$. The upper panels show the even- and odd-parity responses of the gapless wire using $\VQD^*=-0.530\meV$; the two responses occupy distinct narrow gate intervals. The lower panels show the corresponding topological responses using $\VQD^*=0.636\meV$; both occupy a common, substantially broader gate region centered near $\dVQD=0$. Cyan dashed lines mark $\dVQD=0$.}
\end{figure*}

We first vary the dot potential while keeping $\mu=-0.480\meV$ fixed. As shown in Table~\ref{tab:VQD_sector_weights}, the distribution among the particle-number sectors changes rapidly across the displayed gate interval. At $\dVQD=-50\ueV$, the total even-parity probability is approximately $88.10\%$ and is dominated by the $N=2$ sector. At $\dVQD=+50\ueV$, the total odd-parity probability is approximately $88.62\%$ and is dominated by the $N=1$ sector. The $N=0$ probability also increases to approximately $11\%$ at positive detuning, showing that the empty sector becomes relevant in this gate region. In the unrestricted ground-state calculation of Fig.~\ref{fig:unrestricted_deltaC}, the $N=0$ sector becomes the lowest-energy even sector and replaces the $N=2$ state, removing the even-parity capacitance response resonance near $\dVQD=+19.4~\ueV$. Consequently, the balance required for the simultaneous visibility of the two parity branches in the gapless-wire quantum capacitance response does not persist under a sweep of the dot potential.

We next vary the chemical potential while shifting the dot potential oppositely, $\mu=\mu_0+\delta\mu$ and $\VQD=\VQD^*-\delta\mu$, with $\mu_0=-0.480\meV$ and $\VQD^*=0.585\meV$. This compensation approximately preserves the relative dot-wire detuning and therefore gives the two-level gapless wire construction a particularly favorable test of the chemical potential sweep discussed in the main text. It does not, however, preserve the energetic ordering of the many-body sectors. Table~\ref{tab:compensated_mu_sector_weights} shows that the $N=0$ sector carries $99.99\%$ of the probability at $\delta\mu=-50\ueV$, while the $N=3$ sector carries $99.57\%$ at $\delta\mu=+50\ueV$. The $N=1$ and $N=2$ sectors responsible for the Majorana-like response therefore cease to control the equilibrium distribution at sufficiently large detuning even after $\VQD$ is adjusted to compensate the change in $\mu$. Maintaining those sectors would require an additional charging-energy and electrostatic fine-tuning.

The fixed-$N=1,2$ treatment in the main text therefore represents a favorable restriction on the gapless normal-wire capacitance response. Once the accessible particle-number sectors are selected by their energies or sampled thermally, competing sectors can suppress or eliminate one of the two parity branches required for the Majorana-like response. The gapless wire construction consequently requires not only positioning the QD between the two relevant parity-resolved resonances, but also an electrostatic environment that keeps the adjacent $N=1$ and $N=2$ charge sectors accessible as the experimental control parameters are varied.

\section{Rashba spin-orbit coupling and disorder}
\label{app:disorder}

We finally examine whether the distinction identified in the clean systems persists after including Rashba spin-orbit coupling and spatial disorder. We use $\alpha=1.4\meV$ and one fixed disorder realization with root-mean-square strength $V_0=1.2\meV$ and correlation length $\xi_{\rm dis}=20~\mathrm{nm}$. The remaining parameters are specified in Sec.~\ref{sec:model}. The analogous gapless wire configuration is constructed using the same semiconductor parameters and disorder profile as the superconducting wire, but with the proximity-induced pairing removed. The results presented here correspond to this representative disorder realization and are not disorder averaged.

We select the superconducting operating point using the signed topological stability indicator
\begin{equation}
T_s=Q_{\rm Pf}
\frac{\min\!\left[E_g(0),E_g(\pi)\right]}
{\max\!\left[E_g(0),E_g(\pi)\right]},
\label{eq:Ts}
\end{equation}
where $Q_{\rm Pf}=\pm1$ is the real-space Pfaffian invariant and $E_g(0)$ and $E_g(\pi)$ are the excitation gaps obtained using periodic and antiperiodic boundary conditions, respectively. In our convention, $T_s<0$ identifies the topological sector. The magnitude $|T_s|$ measures the relative stability of the finite-size excitation gap under a change of boundary conditions. Fig.~\ref{fig:Ts_map} shows the resulting stability map. The star marks the point $(\Gamma,\mu)=(0.60,1.58)\meV$, which lies inside a finite topological island with $T_s<0$ and is used for the comparison below.

At this common semiconductor operating point, the resonant QD potentials for the gapless wire and topological wire calculations occur at different absolute values. We therefore use the reference potentials $\VQD^*=-0.530\meV$ for the gapless wire and $\VQD^*=0.636\meV$ for the topological wire, defining $\dVQD=\VQD-\VQD^*$ separately in each case. The upper row of Fig.~\ref{fig:realistic_maps} shows the parity-resolved response of the gapless wire. The even- and odd-parity features occupy distinct narrow dot-potential intervals on opposite sides of the reference $\VQD$. At this representative point, they do not exhibit the complementary flux modulation found in the fine tuned clean wire results, indicating that reproducing the Majorana-like response would require further tuning of the other system parameters. In the topological wire, shown in the lower row, both parity responses are centered near $\dVQD=0$ and extend over essentially the same broad gate interval. Consequently, moving the QD potential toward the resonant value associated with either parity sector of the gapless wire moves it away from the resonant value associated with the other, whereas both topological parity sectors remain within a common capacitance-response region.

\begin{figure}[t]
\centering
\includegraphics[width=\columnwidth]{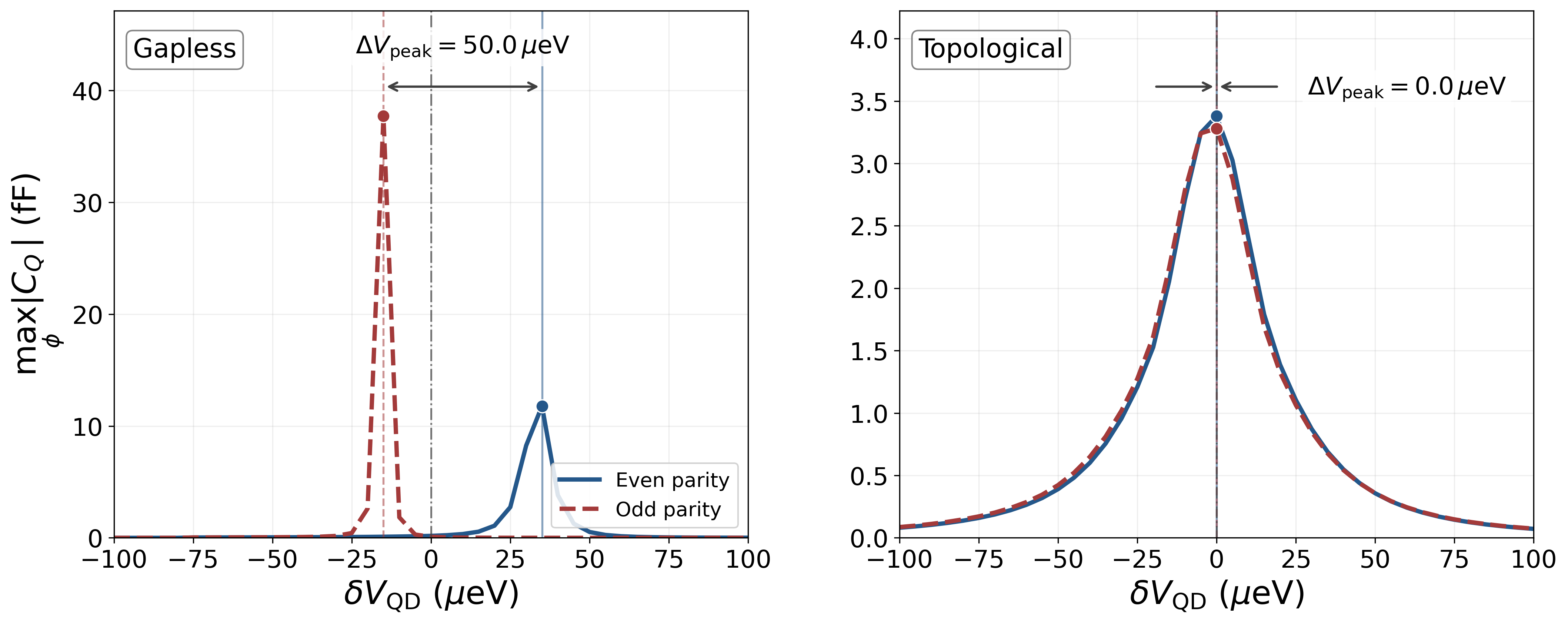}
\caption{\label{fig:realistic_peaks}
\textbf{Peak-capacitance comparison with Rashba coupling and disorder.}
Maximum of $|\CQ|$ over flux as a function of $\dVQD$ for the gapless wire (left) and topological wire (right) at the operating point marked in Fig.~\ref{fig:Ts_map}. Solid blue and dashed red curves denote even and odd parity, respectively. The gapless wire maxima occur near $+35\ueV$ and $-15\ueV$ and are separated by $50.0\ueV$, whereas the topological wire maxima coincide on the numerical gate grid and belong to the same broad resonance envelope. Colored vertical lines mark the individual maxima, the black dash-dotted line marks $\dVQD=0$.}
\end{figure}

Fig.~\ref{fig:realistic_peaks} quantifies this distinction using the maximum of $|\CQ|$ over flux. In the gapless wire, the odd- and even-parity maxima occur near $\dVQD=-15\ueV$ and $+35\ueV$, respectively, giving a separation of $50.0\ueV$. The two peak magnitudes are also strongly asymmetric. In the topological wire, the two maxima coincide at $\dVQD=0$ on the numerical gate grid, and their profiles remain nearly indistinguishable throughout the displayed interval. This numerical coincidence does not require the Majorana splitting to vanish; it implies only that the corresponding displacement of the two capacitance maxima is smaller than the dote potential resolution at this operating point ($<0.1~\ueV$).

Because the clean and disordered calculations use different values of $\mu$ and $\Gamma$, the change in peak separation should not be attributed to disorder alone. The main conclusion is instead that, for the fixed disorder realization considered here, the distinction identified in the clean systems survives: the topological parity responses remain centered within a common broad gate region, while the parity responses of the gapless normal wire segment remain tied to two separated and asymmetric resonant QD-potential values. Rashba coupling and disorder therefore do not rule out an accidental false positive from the gapless normal wire segment, and Ref.~\cite{PinchenkovaKozinHunenbergerLossKlinovaja2026} shows that such a response can be recovered by appropriate retuning. However, simply removing the proximity-induced superconductivity at the same disorder realization and semiconductor operating point does not generate the required alignment. Realizing the Majorana-like response in the gapless normal wire segment still requires additional fine-tuning of the relevant level energies, dot potential, and end couplings.

\bibliography{bibliography}

\end{document}